# Insights into brain microstructure from in vivo DW-MRS

Marco Palombo[a]*, Noam Shemesh[b], Itamar Ronen[c], Julien Valette[d,e]

**Affiliations:**

[a] Department of Computer Science and Centre for Medical Image Computing, University College of London, Gower Street, London WC1E 6BT, United Kingdom.

[b] Champalimaud Neuroscience Programme, Champalimaud Centre for the Unknown, Av. Brasilia 1400-038, Lisbon, Portugal.

[c] C. J. Gorter Center for High Field MRI, Department of Radiology, Leiden University Medical Center, Leiden 2333ZA, The Netherlands

[d] Commissariat à l'Energie Atomique et aux Energies Alternatives (CEA), Direction de la Recherche Fondamentale (DRF), Institut de Biologie François Jacob, MIRCen, F-92260 Fontenay-aux-Roses, France.

[e] Centre National de la Recherche Scientifique (CNRS), Université Paris-Sud, Université Paris-Saclay, UMR 9199, Neurodegenerative Diseases Laboratory, F-92260 Fontenay-aux-Roses, France

***Corresponding Author:*** Dr. Marco Palombo, Department of Computer Science and Centre for Medical Image Computing, University College of London, Gower Street, London WC1E 6BT, United Kingdom. E-mail: mrc.palombo@gmail.com.

**Abstract**

Many developmental processes, such as plasticity and aging, or pathological processes such as neurological diseases are characterized by modulations of specific cellular types and their microstructures. Diffusion-weighted Magnetic Resonance Imaging (DW-MRI) is a powerful technique for probing microstructure, yet its information arises from the ubiquitous, non-specific water signal. By contrast, diffusion-weighted Magnetic Resonance Spectroscopy (DW-MRS) allows specific characterizations of tissues such as brain and muscle *in vivo* by quantifying the diffusion properties of MR-observable metabolites. Many brain metabolites are predominantly intracellular, and some of them are preferentially localized in specific brain cell populations, e.g., neurons and glia. Given the microstructural sensitivity of diffusion-encoding filters, investigation of metabolite diffusion properties using DW-MRS can thus provide exclusive cell and compartment-specific information. Furthermore, since many models and assumptions are used for quantification of water diffusion, metabolite diffusion may serve to generate *a-priori* information for model selection in DW-MRI. However, DW-MRS measurements are extremely challenging, from the acquisition to the accurate and correct analysis and quantification stages. In this review, we survey the state-of-the-art methods that have been developed for the robust acquisition, quantification and analysis of DW-MRS data and discuss the potential relevance of DW-MRS for elucidating brain microstructure *in vivo*. The review highlights that when accurate data on the diffusion of multiple metabolites is combined with accurate computational and geometrical modelling, DW-MRS can provide unique cell-specific information on the intracellular structure of brain tissue, in health and disease, which could serve as incentives for further application *in vivo* in human research and clinical MRI.

***Highlights:***

- diffusion-weighted MRS (DW-MRS) allows to investigate brain metabolite diffusion
- most brain metabolites are predominantly intracellular and cell specific
- diffusion properties of brain metabolites provide exclusive cell-specific information
- the most recent methods for DW-MRS data acquisition and analysis are reviewed
- the potential relevance of DW-MRS for elucidating brain microstructure is discussed.

## Introduction

Nuclear Magnetic Resonance (NMR) signals are incredibly rich sources of information because they are sensitive to a multitude of physical mechanisms, each capable of portraying different physicochemical aspects of the spin ensembles. The chemical shift – the slight, ppm-scale variations in NMR frequency due to electronic interactions in a given molecule – was perhaps one of the earliest NMR interactions to be discovered, and its utilization has transformed Chemistry due to its ability to characterize molecular structure, first in small molecules and then in much bigger molecules such as proteins and membranes. It is therefore not a surprise that NMR's spatially-localized version – rebranded Magnetic Resonance Spectroscopy (MRS) – has had a strong impact on biomedicine and neuroscience. The spectral dimension revealed by MRS in general, and by $^{1}$H MRS in particular, allows a unique way to characterize tissues *in vivo* by quantifying the metabolite levels, at least for those that exist at relatively high concentrations (typically > 1mM). In high-quality MRS spectra, about 20 metabolites have been identified and quantified *in vivo*, including neurotransmitters, energy-related metabolites, osmolytes and others (1). Accurate information on metabolite concentrations in tissue found vast applications: to name but a few, MRS has been used for characterization of energy-cycle metabolism *in vivo* (2-4), identifying and differentiating between tumor types (5) and gaining insights into neurotransmitter turnover upon activation (6). Such studies have been instrumental for both clinical applications as well as for basic science.

Diffusion is another mechanism that has made a huge impact in biomedicine, neuroscience and other fields. The spins giving rise to sharp NMR signals are never stationary, and in the presence of magnetic field gradients – be they internal, i.e. caused by tissue susceptibility variations (7), or externally applied (8) – the NMR signal will also reflect information on the diffusion process that the ensemble had undergone. In solution, the signal attenuation in the presence of magnetic field gradients reflects the diffusion coefficient, which have been used for example, as a second dimension for resolving the constituents in chemical mixtures in which chemical shift spectral lines overlap (9). When diffusion is restricted by the presence of physical barriers, the diffusion signal becomes imprinted with signatures of the confining geometry, and hence its measurement may be capable of reporting on microstructure.

Fortuitously, the typical time-scales of diffusion encodings are compatible with the time required for water molecules to traverse (sub)cellular length scales at 37 °C. Thus, water diffusion is sensitive to structures on a spatial scale much smaller than the voxel size, an advantage that found applications abound, for example, in early detection of stroke (10-12) as well as in fiber orientation mapping and tracking (13, 14) as well as other applications in biomedicine (15).

Why then would we go beyond quantifying microstructures from water signals? Water is present in almost every microscopic as well as macroscopic tissue subcomponent, and is thus inherently non-specific. Water in blood vessels, CSF and cysts diffuses with a very high diffusion coefficient, and flow is also present. In neural tissue, all cells, including astrocytes, neurons, oligodendrocytes or other glia, contain water; furthermore, their subcellular units (cell body, neurites/astrocytic processes, cell mitochondria or cytoplasm, etc.) are also very similar in water content. Different cell types such as excitatory/inhibitory/modulatory neurons, are also not different from the water content perspective. In addition to water in cells, it is important to note also the extracellular space, which also contains water protons at similar concentration (~110M) as the intracellular environment. Since the exact diffusion coefficients or properties in all cellular, subcellular, and extracellular compartments are still widely unknown (16, 17), it is extremely difficult to gain compartment-specific information from water-based measurements.

By contrast with water, most metabolites are predominantly intracellular, and some metabolites are even cell-specific; that is, they are produced by or confined to a specific cellular population (**Figure 1**). Perhaps the most obvious example is N-Acetylaspartate (NAA): numerous studies (18-20) have found that in the nervous system, the osmolyte NAA is produced mainly by neurons, and furthermore, NAA is not secreted by neurons. A very recent study evidenced the production of NAA by oligodendrocytes and confinement of that NAA pool in the myelin that unsheathes the neuronal axons (21). Hence, NAA is considered specific to the neuronal (or neuronal-associated) intracellular compartment. Luckily, NAA is also one of the most quantifiable metabolites, having both very high concentration (~20mM), higher than all other metabolites other than glutamate (Glu), and a very convenient singlet at the 2.02 ppm resonance, which is not modulated over different TEs. In some cases, there are some modulated metabolite signals that can “hide” under NAA’s resonance, especially at lower fields and/or when lines are somewhat broad (e.g. a 1.95ppm GABA methine quintet).

Another metabolite with high compartmental specificity is myo-inositol (Ins). In the CNS, Ins has been shown to be predominantly localized in astrocytes (18, 22, 23). This provides, in principle, a very elegant counterpart to NAA's neuronal preponderance, as Ins represents the complementary major intracellular compartment in the CNS. Other metabolites, such as the co-measured pair creatine and phosphocreatine (tCr) (24, 25) and the co-measured choline compounds (tCho), have been shown to originate predominantly from Glia (up to ~75% specificity for tCho) (23). Le Belle et al. (26) reported tCho concentration of 1.5-2.2 µmol/g in brain tissue using two different extraction techniques, and 27-41 µmol/g in astrocytes. When normalized to tCr, this corresponds to [tCho]/[tCr]=0.15-0.25 in brain tissue (close to what is indeed typically measured in the brain *in vivo* (4)), and [tCho]/[tCr]=1.9-2.4 in astrocytes. This represents at least a 10 times higher concentration in astrocytes than in neurons. Though maybe less-specific than the Ins counterpart, the signal of tCr and tCho is readily quantifiable as both represent uncoupled N-bound methyl singlets.
The motivation for quantifying diffusion properties of these metabolites is now perhaps clearer: their diffusion properties may reflect specific cell-type geometry. Since many diseases or disease phases are characterized by injury to specific cellular types (e.g., glioblastoma affecting glial cells) or compartments, and since water signals are not necessarily representative of any specific compartment, metabolite signals can serve as biomarkers with enhanced specificity. Additionally, information on how metabolites such as neurotransmitters change compartments (e.g., upon neurotransmission) could complement fMRI with a more direct observation of neural activation *in vivo* (27, 28). Furthermore, since many models and assumptions are used for quantification of water diffusion, metabolite diffusion may serve to generate *a-priori* information for model selection.

The purpose of this review is to introduce the potential relevance of diffusion-weighted MRS for elucidating brain microstructure *in vivo*. For a more exhaustive survey of MRS or diffusion at large, the reader is referred to the following exhaustive reviews of diffusion-weighted MRS (29-31), and DW-MRI in general (17, 32-35).

For this review, the reader is assumed to be familiar with general definitions and notations of molecular diffusion and NMR diffusion measurements, and in particular with DW-MRI.

The review is structured as follows: in **Section 1**, methodological aspects of data acquisition and quantification in DW-MRS are introduced and compared to DW-MRI; in **Section 2**, the most recent models proposed to describe diffusion of brain metabolites in the intra-cellular space are reviewed, together with some examples of recent successful applications to experimental data based on advanced diffusion encoding techniques; In **Section 3**, we briefly discuss co-analysis of DW-MRI and DW-MRS data: what has been done so far and what are some of the possible directions in which this approach can evolve to provide a more comprehensive picture of tissue microstructure.

## 1. DW-MRS methodology

In this section, we present methodological aspects of data acquisition and quantification specific to DW-MRS, as compared to DW-MRI or non-diffusion weighted MRS. For more exhaustive reviews of DW-MRS basic methods and applications, the reader is referred to (29-31).

Sometimes, NMR jargon is used. For those readers who are not sufficiently familiar with that jargon, we refer to (36, 37) for help.

### 1.1. DW-MRS pulse sequences: specific constraints, traditional approaches, recent developments

*In vivo* DW-MRS pulse sequences must achieve several goals simultaneously: (1) adequate diffusion-weighting, (2) efficient and accurate signal localization to the volume of interest and (3) robust acquisition of free induction decay (FID) to obtain the chemical shift spectral dimension. This is usually achieved by implementing diffusion-sensitizing gradients within a conventional localized MRS sequence, such as STEAM (STimulated Echo Acquisition Mode) (38), PRESS (Point RESolved Spectroscopy) (39) or LASER (Localization by Adiabatic SlicE Refocusing) (40).

In the early days of DW-MRS, because of the relatively low maximal gradient strength available on MRI scanners, the stimulated echo-based DW-STEAM pulse sequence (41, 42) was the most convenient way to reach sufficiently high b-values *in vivo* (typically 5-10 times stronger than that used for water, since metabolite apparent diffusion coefficient, or ADC, is about 0.1-0.2 µm²/ms versus 0.5-1 µm²/ms for water in tissue, suggesting that for the product bD to attenuate, larger b-values are required for metabolite signals (43)). Towards this goal, the diffusion time is built around the mixing time, which can be long without excessive signal loss due to relaxation (as during the mixing time magnetization relaxes according to $T_1$, typically 1-2 seconds for metabolites (44, 45)): diffusion-sensitizing gradient lobes are inserted within the echo time, between the first and second slice-selective 90° RF pulses for dephasing, and after the third slice-selective 90° RF pulse for rephasing (see **Figure 2A**). This is advantageous to reach high b-values but also to specifically study diffusion at long diffusion times. In addition, because of the short echo time that can be reached, STEAM allows quantifying the signal of more heavily J-modulated metabolites such as Glu or Ins. Another noteworthy feature of STEAM is its generally excellent performance in terms of water suppression, because an additional water suppression RF pulse can be inserted during the mixing time. However, one issue with STEAM is that the cross-terms with other gradients tend to be very large, as shown in **Figure 2B**. This is because cross-terms build up during the mixing time, which can be quite long in STEAM, and furthermore, the cross-terms are proportional to the cube of time, making these effects potentially highly significant. It is therefore critical to pay extra attention to cross-terms, either by calculating them, or minimizing their effect by performing two successive acquisitions with opposite diffusion gradient polarities and calculating the geometric average of signal attenuations measured with both polarities (46-49). The most significant drawback with stimulated echo results is certainly the loss of half the magnetization. Therefore, a spin echo sequence might be preferred when experimentally possible, as sensitivity is one of the major issues of MRS in general, and of DW-MRS in particular.

The first spin echo DW-MRS sequences were based on PRESS, and performed on preclinical systems as gradient strength was enough to reach sufficient b-values (50, 51). Identical pairs of gradient pulses G were inserted around one or the two slice-selective 180° RF pulses (90°; G; 180°; G; G; 180°; G), which allowed compensating for phase variation due to bulk translational motion, provided that motion remains constant during the sequence. The sequence was subsequently modified for clinical implementation by applying a bipolar gradient scheme (52) (90°; G; 180°; -G; G; 180°; -G) (**Figure 2C**), thus taking advantage of the twice-refocused spin echo to increase maximal achievable diffusion-weighting, while minimizing eddy currents and mitigating, to a certain extent, cross-terms with time-constant background gradients (49, 53). A DW-PRESS sequence with specific gradient scheme has also been proposed to provide single-shot isotropic diffusion-weighting (i.e. weighting by the diffusion tensor) (54), which may be of interest if one is only interested in rotationally invariant indexes, e.g. the mean diffusivity, in highly anisotropic voxels such as in white matter bundles.

Another type of double spin-echo sequence incorporating fully adiabatic refocusing pulses, termed LASER (40), has been gaining increasing attention. LASER offers superior localization performance compared to STEAM or PRESS, as well as insensitivity to inhomogeneities of the radiofrequency making LASER a particularly attractive localization scheme, also for diffusion-weighted MRS as pioneered in (55). In one variant, oscillating gradients were inserted around the first 180° refocusing pulse to measure metabolite diffusion at very short time scales (56, 57). In another variant, three pairs of gradient pulses of opposite polarities were inserted around the three pairs of refocusing pulses to achieve single-shot isotropic diffusion-weighting while minimizing cross-terms with other gradients (58). In addition to previously mentioned advantages, the six successive refocusing pulses in LASER make it a CPMG sequence, resulting in slightly longer $T_2$ and reduced J-modulation compared to PRESS, which is beneficial for the detection of Glu and Ins. Unfortunately, LASER may be difficult to perform in Humans at high fields (≥7 T) because of specific absorption rate (SAR) issues, but more practical variants such as semi-LASER, in which the excitation pulse is a non-adiabatic selective pulse, may be employed.

A recent development in MRS sequence design consists in temporally isolating diffusion-weighting from localization. So far, this strategy has been implemented on preclinical systems by successively playing out three blocks: non-spatially-selective excitation, diffusion-weighting, and LASER localization (59, 60), as shown in **Figure 2D** and **Figure 3**. The main advantages are the absence of cross-terms between diffusion and localization gradients, a very clean, self-refocused LASER localization, and greater flexibility of the sequence. Using such a block-based scheme, the diffusion block can be readily designed according to specific experimental questions, independently of the localization block. For example, a Double Diffusion Encoding (DDE) scheme could be implemented using this approach (59), giving rise to excellent SNR that enabled measurements of metabolite microscopic anisotropy (**Figure 3**); a stimulated-echo scheme was implemented in (60) to reach very high b values and study how metabolite signal attenuation depended on TE and TM without any bias due to cross-terms (**Figure 2D**). Shemesh et al. exploited spectrally-selective excitation and refocusing to precisely control the spectral profile, manipulating only metabolites/chemical shifts of interest independently of the localization itself. Spectra acquired using this approach were termed Relaxation Enhanced MRS (RE MRS), as they benefit from the absence of water signal and the potential gain in signal enhancements due to relaxation enhancement arising from potential cross relaxation effects between metabolites and water. Adapting this approach to clinical systems remains to be done and might be difficult with a LASER localization block (because of the SAR and of the relatively long TE that the block design would impose), but an ISIS (Image-Selected In vivo Spectroscopy (61)) localization block may be used instead to overcome these limitations. It is important to mention that a small diffusion-weighting is generated by the localization block (~0.5 ms/µm², for example see (60). However, unlike cross-terms, this constant b term disappears when taking the log of signal attenuation, e.g. when measuring the ADC. In fact, this is one of the advantages of using such separated approaches. However, if one prefers, it can also be readily taken into account for computation of the exact b-value, or the b-value generated by the diffusion block can be reduced accordingly.

Whatever the sequence used, an important constraint arising from the low concentration of metabolites (typically ~10,000 times less) than water, is that voxel size and acquisition time generally need to be increased compared to DW-MRI, which is a major hurdle to the applicability of DW-MRS. In the rodent brain for example, using high field scanners and state-of-the-art RF coils, acquiring a single b-value along a single direction requires a few minutes to get decent SNR value in a voxel of a few µL (i.e. corresponding to a given cerebral structure, such as the hippocampus or the striatum), and acquisition time needs to be increased up to ~1 hour to acquire one spectrum at very long diffusion time (longer than the relaxation time). In Humans, where acquisition time is critical, temporal resolutions of a few minutes are compatible with spatial resolutions of a few cubic centimeters. For example, diffusion tensor MRS was performed on a 3 T scanner equipped with 20 mT/m gradients, in ~10 $cm^3$ voxels in the Human brain, using six directions with two b-values, within ~40 minutes using a STEAM sequence (62) and a PRESS sequence (43). More recently, a PRESS-based diffusion tensor MRS protocol was performed on a 7 T scanner equipped with 40 mT/m gradients in a 3.6 $cm^3$ voxel in Human corpus callosum, using six directions with two b-values in addition to the reference spectrum at b=0, within 15 minutes (63).

## 1.2. Post-processing: from raw data to unbiased signal attenuation

Compared to DW-MRI, additional sources of bias can lead to incorrect quantification of DW-MRS signal attenuation. Fortunately, most of these sources of bias can be accounted for, or at least partially controlled. Key post-processing steps to obtain "unbiased" signal attenuation have already been detailed in a recent review (31), but will nonetheless be briefly reviewed here, as unbiased measurement is critical in the perspective of microstructure modelling.

Bulk translational motion in the presence of diffusion gradients results in signal phase variations. This is problematic in MRS, because many scans are generally averaged to get sufficient SNR, and averaging of spectra with different phases will result in overall signal attenuation and/or distortion. If enough metabolite signal can be detected in a single transient, phase correction performed directly on each transient prior to averaging has been shown to be effective to overcome this artifact (42). Alternatively, if metabolite signal is too low, residual water signals may be used, or the full water signal when metabolite cycling is performed (64) (65). However, care should be taken when using the water signal as a phase correction reference, as it does not necessarily originate from the same compartment, in particular if the voxel contains CSF. In such cases, phase correction based on water may not fully restore metabolite phase. To circumvent this problem, it has been proposed to use the water signal after adding an inversion-recovery CSF-nulling block (64). Other kinds of bulk motion, such as rotational motion, result in overall signal attenuation on individual scans, and are therefore less trivial to correct. If sufficient signal is available on single scans, it is possible to select and discard scans exhibiting abnormal signal attenuation e.g., below a certain threshold determined from the highest signal outcome of the series and taking SNR into account. Macromolecule signal at 0.9 ppm, which does not overlap with metabolite signal, has been used in a recent study using oscillating gradients (57) as a reference of approximately null diffusivity, at least at low b-values, so that scans with decreased MM signal compared with b=0 could be discarded as presumably resulting from bulk motion artefact.

The effect of eddy currents on spectral distortion has been shown to be relatively easy to correct for (61), provided a water reference signal was acquired using the same sequence, i.e. data are acquired with the same volume selection, same sequence timing and the same diffusion-weighting, with water-suppression RF pulses turned off. Assuming only the water resonance contributes to this reference signal (which is generally the case when no water suppression is applied and diffusion-weighting is not too high), the temporal phase distortion induced by eddy current can be directly assessed by measuring the phase $\phi_{EC}(t)$ of the free induction decay. Then, eddy currents correction on the water-suppressed free induction decay $FID_{metab}(t)$ is achieved by simply removing the eddy current-induced phase, i.e. by computing $FID_{metab,ECC}(t)=FID_{metab}(t)/\exp(i\phi_{EC}(t))$. This approach is compromised in situations where metabolite signal significantly contributes to the reference signal acquired without water suppression, for example at very high b values. In that case, a linear-prediction singular value decomposition (LPSVD) algorithm allows isolating the water signal before calculating $\phi_{EC}(t)$ (66). This approach may also be useful in poor SNR conditions.

Once MRS spectra have been properly reconstructed, metabolite signal needs to be quantified. This was historically done by measuring peak height, area, or ratios, which can be accurate when the baseline is flat and when the different metabolites do not overlap. However, to measure diffusion properties for as many metabolites as possible, including J-modulated metabolite such as Glu and Ins, short TE sequences are required, resulting in many overlapping multiplets as well as strong macromolecule contribution (see below). In such cases, it is much more reliable to analyze spectra in terms of linear combination of individual metabolite signals (**Figure 4**), as performed, e.g., by LCModel (67). Other analysis software exist, such as AMARES (68), TARQUIN (69), FID-A (www.github.com/CIC-methods/FID-A), and these also quantify spectra in similar ways. This permits the utilization of the exhaustive spectral information, in particular by simultaneously fitting peaks at different chemical shifts but belonging to the same metabolite. In the end, it is critical to obtain very robust signal quantification, with more stringent quality criteria than for standard MRS. For example, when using LCModel or TARQUIN, quantification precision may be evaluated based on the Cramér-Rao lower bounds (CRLB). CRLB is an estimation of the standard deviation of the measurement. When measuring the ADC with a two-point experiment (e.g. b=0 and b=2 ms/µm²), error propagation analysis yields the following expression for the standard deviation on ADC, as expressed in terms of CRLB provided by LCModel:

$$\sigma_{ADC} \sim \frac{1}{b} \times \sqrt{{CRLB_{b=2}}^2 + {CRLB_{b=0}}^2} \qquad [1]$$

Assuming $CRLB_{b=2}$~$CRLB_{b=0}$~20%, often considered as a reliability threshold in the MRS literature, we get s.d.(ADC)~0.14 µm²/ms, which is comparable to a typical ADC. In contrast, a more demanding quality threshold, i.e. CRLB~5%, will result in s.d.(ADC)~0.035 µm²/ms, which is well below typical ADC.

A very important and often overlooked issue related to metabolite quantification is the macromolecule (MM) signal. MM consist of large molecules, e.g. proteins, with relatively short $T_2$ (a few tens of ms) resulting in broad resonances overlapping with metabolites (70). Not surprisingly, MM diffusion has been reported to be very slow, with ADC in the brain found to be ~0.005 to ~0.01 µm²/ms and a close to mono-exponential attenuation even at very large b-values (60, 71). MM signal is not negligible compared to metabolites when relatively short TE sequences are used (typically less than 50 ms), and it has been shown to significantly affect estimated metabolite diffusion if not properly accounted for (72), especially at higher b-values where MM have a higher relative signal due to their slow diffusion (**Figure 4**). In this context, it is absolutely critical to carefully disentangle metabolite from MM signal. The spectral quantification software LCModel (67) offers the possibility to include a group of independent broad contributions to model MM signal, but this will generally introduce too much bias and variability, as overlapping MM and metabolite signal can still be confounded. The method of choice is instead to acquire an experimental MM spectrum using metabolite-nulling acquisition (73) and to incorporate this spectrum in LCModel basis-set. This will ensure that the actual MM contribution is considered, thus limiting the variability in the estimation of the MM signal at different b values, in particular because the isolated MM peak at 0.9 ppm will result in reliable MM estimation.

## 1.3 Early measurements of metabolite DW-MRS signal decays, ADC and Diffusion Tensor MRS

Somewhat similarly to the history of water-based DWI, the initial focus in DW-MRS was in characterizing the signal attenuation, and measuring metabolite ADCs at a given set of diffusion time and gradient amplitude combination using Single Diffusion Encoding (SDE) approaches (74). In perfused cells, van Zijl et al already identified non-exponential decays against b-value (75), while Moonen et al measured metabolite ADCs in muscle tissue (41). Several metabolite ADCs were subsequently measured in the healthy human brain (42, 43, 62, 76) as well as in the in-vivo rat brain (77, 78), reporting ADC values of the three main metabolites: NAA, tCr and tCho, consistent between human and rat brain and around 0.15-0.20 $\mu m^2/ms$. For a summary table of ADC values in various brain regions and animal species, we refer to the comprehensive Table 1 in (79).

Wick et al (80) monitored changes in metabolite ADCs upon ischemia in the rat, and suggested that specific neuronal and astrocytic swellings could be identified from the NAA and Ins signals, respectively. Van der Toor et al (50) measured a significant decrease in NAA and tCr ADCs of 29% and 19%, respectively, compared to the contralateral region, after 3 hours of ischemia in rat brain. Pfeuffer et al (71) were able to use DW-MRS derived ADCs of glucose, lactate and compare them to purely intracellular metabolites to demonstrate the equal partitioning of glucose and lactate across the extra/intra-cellular compartments. Others studied the ADCs in stroke (65, 81, 82) as well as in tumors (83). These early studies all showed that there is value in measuring metabolite diffusion, yet, most of them did not address tissue microstructure directly, but rather indirectly interpreted ADC variations in terms of possible alterations of microstructure.

In analogy to DTI, it is possible to assess the macroscopic anisotropy of metabolite diffusion in the brain tissue by using a minimum of six diffusion gradient directions. This technique is called diffusion tensor spectroscopy (DT-MRS, also known as DTS), and permits the measurement of the elements of the macroscopic diffusion tensor of typical MRS-visible metabolites, such as NAA, tCr and tCho. DT-MRS has been used to assess the anisotropy of metabolite diffusion in human (62, 84, 85), animal model of frog peripheral nerve (86) and bovine optic nerve (87).

In human brain, elevated fractional anisotropy (FA) values of 0.48–0.72 for NAA (62, 84, 85), 0.56–0.73 for tCr (62) and 0.59–0.74 for tCho (62) were observed in major WM regions, such as the corpus callosum, corticospinal tract and arcuate fasciculus. These tensor measurements agreed with specific measurements of anisotropic diffusion of NAA in the human corpus callosum (52), as well as in animal models of frog peripheral nerve (86) and bovine optic nerve (87). Interestingly, a first examination of cortical gray matter (GM) yielded an unexpected high degree of anisotropy (FA=0.53–0.79) for all three metabolites, similar to the WM FA values (62). In contrast, typical FA values of water in cortical GM are approximately 0.2 (88) or less. A later study by Ellegood et al. (43) showed that the high FA measured in GM was an artefact caused by the variability of the directional ADC measurements (89). The greater directional ADC variability in cortical GM was the result of a poor two-point slope estimation of the signal intensity *versus* b value when using a less than ideal maximum b value of 3-5 ms/$\mu m^2$. Using a maximum b value of ~5 ms/$\mu m^2$, instead of the much lower 1.8 ms/$\mu m^2$ used in (62), Ellegood et al. estimated lower FA values of 0.25, 0.30 and 0.28 for NAA, tCr and tCho, respectively, in human occipital GM, while FA values of 0.47, 0.51 and 0.51 for NAA, tCr and tCho, respectively, in human subcortical WM were consistent with the estimates at maximal b of 1.8 ms/$\mu m^2$ (43). A summary of DT-MRS metrics (DT eigenvalues $\lambda_1$, $\lambda_2$ and $\lambda_3$, mean diffusivity, MD, and FA) estimated in different WM and GM regions in healthy human brain are reported in **Table 1**. The DT-MRS metrics estimated by different research groups with different scanners and different magnetic fields, but similar echo time and diffusion time, show a good consistence and reproducibility (**Table 1**).

## 2. What does DW-MRS data tell about brain cells microstructure?

Metabolite ADC and DT derived metrics, while very useful in some cases, are only indirect reporters of microstructure as they reflect the overall macroscopic reduction in diffusion imposed by geometry in the Gaussian diffusion limit. Earlier strategies for quantifying microstructure from metabolite DW-MRS have involved q-space MRS (90, 91), which clearly showed non-monoexponential diffusion for NAA, and even quantified some of its time-dependent properties by observing the average propagator at different diffusion times. Others proposed a first attempt to model NAA diffusion taking into account the cellular structure (35, 52), while more recently, the ADC time dependence at (ultra)-short td was linked to local cellular geometry such as fiber diameter (56, 57), while the ADC at (ultra-)long $t_d$ was linked to the long-range cell morphology (92).

In this section, the most recent frameworks proposed to relate brain metabolites diffusion in the intra-cellular space to microstructure are reviewed, together with some examples of recent successful applications to experimental data, which support their validity. We focus our interest on geometrical models which link the measured diffusion-sensitized echo signal attenuation and/or derived diffusion metrics, such as ADC, to cellular microstructural determinants, such as fibers diameter and length, number of embranchments and others).

### 2.1 Molecular diffusion and DW-MR signal

Conventional spin-echo and stimulated-echo MRI and MRS can be sensitized to diffusion by symmetrically applying magnetic field gradient pulses that attenuate the echo signal S. In the case of non-restricted or Gaussian diffusion, the effect of molecular diffusion leads to an exponential attenuation of S with the b-value. However, in the general case of diffusion in biological tissue, the presence of restrictive or hindering boundaries (membranes, cytoskeleton, macromolecules, organelles etc…) drastically influences the motion of probe molecules (water or metabolites) and the consequent signal attenuation and the derived ADC. In particular, the echo signal attenuation S is no longer a simple exponential decay and the measured ADC is in this case depends on $t_d$ (34, 93-95).

In the case of water diffusion, the description and interpretation of the measured signal and derived metrics such as the ADC in terms of the underlying tissue microstructure is generally very difficult, due to the non-specificity of the water signal and the complexity of the tissue as a whole. Non-negligible water volume fraction in the extracellular space, whose geometric properties can be very difficult to assess, as well as cell membrane permeability have to be taken into account. Phenomena related to the extracellular space, such as intra-/extra-cellular water exchange, potential flow of the cerebrospinal fluid (CSF), extracellular volume fraction, diffusivity and tortuosity in the extracellular space, cannot be neglected for water signals, but can be neglected when interpreting and modelling intracellular metabolite diffusion. For some metabolites, diffusion properties are expected to depend mostly on intracellular parameters such as cytosol viscosity, molecular crowding and binding, as well as the size and shape of the cellular compartment, and metabolite diffusion modelling is thus much simpler, allowing for a more direct and precise estimate of the specific cellular compartment features.

## 2.2 Modelling metabolite intra-cellular diffusion

It is important to distinguish the case of restricted diffusion from hindered diffusion, because the echo signal attenuation and the ADC diffusion time dependence for these two scenarios are different.

*Hindering effects: tortuosity and obstruction*

Numerous immobile obstacles (i.e. considering the typical time window of DW-MRS experiments) exist within the intracellular space, such as the cytoskeleton and various organelles, potentially making the cytosol a tortuous space. The tortuosity $\tau$ refers to the effect of hindrances imposed by various *immobile* obstacles in the medium on the path that diffusing molecules can take. Such hindrances affect the minimal pathway between two points so that the pathway becomes tortuous, rather than a straight line. In this case, the shortest pathway between two points is increased, on average, by a factor $\tau$ (≥1), compared to a straight line connecting these points. At very short $t_d$, during which the mean square displacement (MSD) is very small compared to the square of the typical distance between obstacles, tortuosity will not affect molecular displacement, and the diffusion process will appear free. At longer $t_d$, in d dimensions, one has (96):

$$MSD \sim 2d \, \frac{D_{free}}{\tau^2} \, t_d \qquad [2]$$

Equivalently, the ADC measured at *long* $t_d$ converges to:

$$\lim_{t_d \to \infty} ADC = ADC_\infty \sim \frac{D_{free}}{\tau^2} \qquad [3]$$

The way in which the ADC approaches its tortuosity limit and the value of the tortuosity itself in Eq. [3] may be affected by secondary structures of cell morphology such as dendritic spines, astrocytic leaflets or axonal beads (97). These secondary structures can be seen as randomly distributed hindering sources to metabolites diffusion along cell fibers. In this case, a well-behaved and specific ADC power law time dependence is expected for mono-dimensional short-range disordered (hindering) structures: $ADC \sim ADC_\infty + C\, t_d^{-0.5}$ (95). Recent numerical simulations and experimental results (97) suggest that structures such as these can also affect metabolite diffusion and the measured echo signal attenuation at high q/b values. Consequently, they have a non-negligible effect on the derived diffusion metrics, such as ADC power law time dependence, and on the estimated cell geometrical parameters such as fiber radius (97).

A related but nevertheless distinct phenomenon compared with tortuosity that can also lead to a decrease in the measured ADC is that of obstruction. In cell cytoplasm, macromolecules cannot be neglected compared to the solvent concentration, thus smaller molecules (e.g., metabolites) have to skirt around the larger and generally irregularly shaped obstructing molecules, increasing their diffusion path length. Obstruction is a complicated many-body problem which is in essence hindrance by a time-dependent geometry - the obstructing molecules are also moving - in which the interactions between the particles need to be considered. Obstruction results in the measured ADC being reduced by a factor that depends on the concentration of the particles, as well as perhaps by other factors, such as electrolyte friction and solvation (96, 98).

The effects leading to obstruction generally operate on very short $t_d$ and length scales and are consequently typically well averaged on experimentally available timescales. Thus, from the perspective of diffusion, obstruction and viscosity are similar in their effects and closely related, and this helps to explain why the measured viscosity, which includes obstruction by macromolecules, can depend on the size of the probe molecule (99).

*Restriction effects*

The confinement of metabolites within a given compartment, such as organelles or even the entire intracellular space, will also impose an upper limit to the displacement variance. While this restriction effect is negligible at very *short* $t_d$, as the displacement variance is very small compared to the square of the typical distance between diffusion barriers, at *long* $t_d$, restriction will strongly influence the diffusion process. For example, in a situation of perfect restriction, i.e., without a possible escape for diffusing molecules, the ADC will converge to 0 with increasing $t_d$, thus, the convergence of the ADC to zero with increasing $t_d$ is a specific signature of restricted diffusion. Similarly, the echo signal attenuation as a function of b deviates from the simple exponential decay. For example, for diffusion within a reflective cylinder of radius *a* and when the diffusion sensitizing gradient is applied along the direction orthogonal to the restricting frontiers, in the short gradient pulses (SGP) approximation (100) the signal attenuation is given by:

$$\frac{S_{cylinder}(q,t_d)}{S(q=0,t_d)} = \frac{[\,2J_1 2\pi qa]^2}{(2\pi qa)^2} + 8(2\pi qa)^2 \sum_{n=1}^{\infty} \left\{ \frac{1}{1+\delta_{n0}} [J_n' 2\pi qa]^2 \sum_{m=1}^{\infty} \frac{\alpha_{nm}^2}{(\alpha_{nm}^2 - n^2)\left[\alpha_{nm}^2 - (2\pi qa)^2\right]^2} e^{-\frac{D_{free}\alpha_{nm}^2 t_d}{a^2}} \right\} \qquad [4]$$

where $J_n$ is the Bessel function of integer order n; $\alpha_{nm}$ is the m-th positive root of the Bessel equation $J'_n = 0$; $\delta_{n0}$ is the Kronecker delta symbol. Note that for other simple restricting geometries like parallel infinite reflective planes or reflective sphere, similar exact analytical solutions have been derived in SGP approximation (101).

As can be inferred from the attenuation behaviour in Eq. [4], as $t_d$ is such that the diffusing spins interact with the enclosing geometry (i.e., $t_d \sim a^2/D_{free}$ where a is the size of the confining geometry) the attenuation profile differs significantly from that of the free diffusion model. When the interactions with the boundary become significant ($t_d \gtrsim a^2/D_{free}$) an interesting effect is noted if the attenuation is plotted as a function of q: diffusive diffraction-like effects arise and structural information about the enclosing geometry can be obtained from the characteristics of the diffraction pattern. For example, the diffractive minima occur at q = n/(2a) (n=1, 2, 3, …). In the early 1990s, Callaghan and Coy proposed for the first time the analogy between NMR measurements in a field gradient and diffraction, formulating its link to the underpinning confining geometry (102, 103). Many subsequent DW-MRI and DW-MRS studies of water and metabolite diffusion in biological tissues showed the potential of this approach to characterize tissue microstructure (87, 104, 105). For a comprehensive review on this topic, we refer the reader to (106).

For the case of cylindrical restrictions, it is not always easy or even possible to apply the diffusion sensitizing gradient exactly along the direction orthogonal to the restricting frontiers. In this case, an expression for the echo signal attenuation in the generic case where the diffusion gradient is applied along a direction separated by an angle θ relative to the cylinder axis is provided by (107):

$$\frac{S^{\theta}_{cylinder}(q,t_d)}{S(q=0,t_d)} = \frac{S_{cylinder}(q_{\perp},t_d)}{S(q=0,t_d)} e^{-D_{free} q_{\|}^2 t_d} \quad [5]$$

where $q_{\perp} = qsin(\theta); q_{\|} = qcos(\theta)$.

Note that analytical equations also exist to calculate the dispersive diffusivity D($\omega$) (i.e. the Fourier Transform of the velocity autocorrelation function) in cylinders and spheres when using oscillating gradients in the low b-value regime (108). However, one of the great advantage of using oscillating gradients is that, at sufficiently high frequencies (i.e. in the Mitra regime, (109)), a model-free linear fit of D($\omega$) as a function of $\omega$ can be done to estimate $D_{free}$ and the surface-to-volume ratio S/V (110). Such a model-free approach might be preferred over geometrical models, provided sufficiently high $\omega$ can be reached.

**2.3 Models for cellular compartments**

Generally speaking, the diffusion of brain metabolites in the intra-cellular space can be modelled as molecular diffusion in the cytosol with restriction mostly due to the cell morphology and internal structure. While the cytosol viscosity, including macromolecular crowding and other effects, can ideally be investigated by performing (ultra-)short $t_d$ experiments where the ADC time dependence at (ultra-)short $t_d$ can be studied without the necessity of any specific modeling (57), the estimation of cell morphology is more complex, and requires more sophisticated modelling.

Recently obtained experimental results, assist the complex modelling of cellular structure. These results suggest that in order to define a proper model for brain cellular compartment, different diffusion regimes and different corresponding models have to be considered, according to the different $t_d$ investigated.

*Cytosol viscosity and macromolecular obstruction*

Measurements performed in conditions where tortuosity and restriction effects are assumed to become small or negligible, i.e., at (ultra-)short $t_d$ (lower than 10 ms) using oscillating gradients (56, 57) yielded values of metabolite $D_{free}$ in the range of ~50% to ~80% of the free diffusivity for those metabolites in aqueous solution. This suggests a cytosolic viscosity (including molecular crowding) that is, at most, twice the value of water.

*Metabolite diffusion primarily occurs in long fibers: a first argument based on ADC time-dependency*

The observed strong decrease in metabolite ADC as $t_d$ is increased from ~1 ms to ~10 ms reported in (56, 57) suggests that metabolite diffusion in brain cells is hindered by obstacles that are typically separated by distances of ≤2 µm. A priori, these obstacles could be either organelles or structures of the cytoskeleton, or simply the membranes of fibers extending from the cell bodies of neurons and glial cells, i.e. axons, dendrites and astrocytic processes.
On the other hand, metabolite ADCs (measured within a spectroscopy voxel containing a mix of white and gray matter) have been shown to be remarkably stable in the mouse and macaque brain in the range of (ultra-)long $t_d$ values between 100 ms and 2 s (48, 92), despite a slight trend to decrease with increasing td. This suggests that metabolites are for the most part not confined inside small subcellular structures, such as organelles or cell bodies, but diffuse along the long fibers characteristic of neurons and astrocytes. Had the metabolites been confined to subcellular structures or to a geometrically closed structure such as a cell body, their ADC($t_d$) would have rapidly approached zero. Metabolite ADC stability has also been confirmed separately in human gray and white matter for $t_d$ between 100 and 720 ms (111).

Thus, brain intracellular metabolites diffusion can be thought primarily as molecular diffusion in cellular processes, described in first approximation as a collection of long cylinders with radius *a* ≤2 μm, and intracellular metabolites diffusivity $D_{free}$ lower than the diffusivity of free metabolites in aqueous solution. The idea of metabolite diffusion being characteristic of diffusion in fibers is consistent with the known cellular morphology of neurons and astrocytes, where long cellular processes (of more than tens, and sometimes up to hundreds of μm) represent most of the cellular volume fraction, about 80% or more (112, 113). Although long axonal fibers may be reasonably considered of infinite length, this assumption may not be valid at all experimentally accessible $t_d$ values for dendrites and astrocytic processes which exhibit a more complex branching structure, as will be discussed later in this review.

*Metabolite diffusion primarily occurs in long fibers: a second argument based on DDE*

In contrast to SDE, multiple diffusion encoding approaches – and in particular, Mitra's DDE approach – can provide insights into microscopic anisotropy – a parameter that disentangles orientation distribution from the "local" anisotropy in a heterogeneous system. For more thorough reviews the reader is referred to (114, 115), however the main notions will be briefly mentioned here. As its name suggests, the DDE sequence (**Figure 3**), first suggested by Cory et al. (116), contains two diffusion sensitizing epochs separated by a mixing time. Mitra was the first to suggest the angular DDE experiment, a variant with potential to disentangle microscopic from macroscopic anisotropy and orientation dispersion from the shape of the curve spanned by varying the relative angle between DDE's gradient pairs (117). Mitra thus suggested an ingenious way to characterize the size and shape of completely disordered systems, which would otherwise appear spherical in SDE experiments: spheres would incur no modulation at long mixing times, whereas randomly oriented but locally anisotropic compartments would give rise to a modulation from which the microscopic anisotropy, a measure of the local pore eccentricity, can be extracted. That is, DDE signals, such as those shown in **Figure 5** from a DDE experiment, can have modulations whose amplitude reflects the microscopic anisotropy – in principle, the deeper these oscillations are (as quantified, e.g., from the signal at 0 and 90 degrees), the larger the microscopic anisotropy in the system. It is important to notice that when anisotropic pores are randomly oriented, the SDE experiment provides isotropic decays from which the microscopic anisotropy or orientation dispersion are nearly impossible to recover. This idea was later theoretically refined (118) and generalized to 3D rotationally invariant schemes (119, 120) and experimentally demonstrated in numerous systems ranging from porous media and phantoms to neural tissue ex- and in-vivo (49, 120-125). The main advantage of such DDE approaches is that they can deliver the information on the microscopic anisotropy directly from only a few points along the angular curve, thereby providing an efficient way of inferring the underlying geometry. As well, they can differentiate between multicomponent Gaussian diffusion and non-Gaussian, restricted diffusion directly from the shape of the angular modulation.

The first spectroscopic DDE implementation utilizing the chemical shift to separate compartments and DDE filters to gain sensitivity towards microstructure was given in Shemesh et al (123) on an emulsion system mimicking a "cellular" and "extracellular" environment. The spectral dimension here was greatly simplified compared to *in vivo* spectra, and contained only two signals: the toluene (intra) and water (extra) signals in roughly equal amounts. SDE q-space experiments could not clearly differentiate the two signals, and in fact it was unclear whether they exhibit compartmentation at all. By contrast, the DDE MRS experiments (non-localized) showed clearly that there are two different diffusion behaviors: the first, representing diffusion within big spheres (toluene) and the second exhibiting restricted diffusion in a randomly oriented environment with a high eccentricity (water). These chemical-shift driven specificity enhancements, coupled with DDE's ability to resolve microscopic anisotropy unambiguously defined the total microstructure of a complex system, and thus provided strong incentives for *in vivo* DDE MRS experiments.

The first *in vivo* DDE MRS experiments were performed using the RE MRS approach specifically tailored for NAA, tCr, tCho, and Lactic acid (Lac) signals in a rat stroke model, where voxels were localized in ipsi- and contra-lateral sides (59). Raw data and results are then shown in **Figure 5B** and **C**. DDE's characteristic modulation curves are easily evident in the raw signal stacked plots, and can be further appreciated from the plots. Several features were noted: (1) the metabolites unequivocally exhibit restricted, non-Gaussian diffusion, and the shape of the curve suggested their localization in highly eccentric compartments; (2) the stroke altered the geometry of the intraneuronal compartment within the voxel, as evident from the statistically significant difference in NAA's modulation between the hemispheres; (3) Lactic acid moves from a rather spherical compartment to a much more eccentric compartment upon ischemia.

A later DDE RE MRS study (126), this time performed in the normal rat brain, used a slightly different excitation/refocusing pulses, such that their bands encompassed only NAA and Ins signals, in an effort to discern between the neuronal and astrocytic compartments more clearly. As mentioned earlier, NAA is a specific intra-neuronal marker, while Ins is present nearly exclusively in the intra-astrocytic compartment. **Figure 6A** shows how the multiband pulse indeed excited only Ins and NAA signals, greatly simplifying the spectrum (n.b., it is displayed in magnitude mode to avoid the deleterious effects of J-coupling and since the spectral resolution is not that important due to the band-limited excitation), and facilitating the DDE acquisition. The clear DDE modulation observed (**Figure 6B**) extends the previous findings, and supports the notion that NAA and Ins are restricted in elongated compartments.

To summarize, the measured characteristic oscillations for NAA, tCr, tCho and later for the Ins signals, can only be attributed to restricted diffusion (**Figure 5** and **6**). Multi-Gaussian diffusion would not give rise to such modulated curves (117). Thus, intracellular diffusion within neurons and astrocytes could be confirmed as non-gaussian from two independent techniques: time dependent SDE, and DDE. Another feature that can be compared between SDE and DDE experiments is the degree of anisotropy in each compartment. DDE's signal modulations revealed clear signatures for microscopic anisotropy, i.e., non-spherical components, with similar yet not identical microstructures (e.g., length/radius) for both neurons and astrocytes. A reasonable hypothesis for this observation is that these experiments represent diffusion in randomly oriented neurites in neurons, whereas in astrocytes, the compartments probed are the randomly oriented astrocytic processes. This is largely consistent with the conclusions drawn for observing ADC time-dependency over a very large range of diffusion times (92).

*Estimating fiber diameter from the high b/q-value signal attenuation*

At intermediate $t_d$, i.e., between 10 and 100 ms, the fraction of metabolites experiencing branching during their diffusion along a cell fiber process may be neglected and the cell fiber can be in good approximation modeled as a long infinite cylinder. Considering an average metabolite $D_{free}$~0.4 μm²/ms and an average fiber length between successive embranchments $L_{segment}$~30 μm (92, 127), for $t_d$≤100 ms the mean metabolite displacement along the cell process is ≤10 μm << $L_{segment}$. In this regime, cellular fibers can be modeled as long hollow cylinders. Because the spectroscopy voxel from which the DW-MRS signal is measured contains a large number of cellular processes randomly oriented in space, a model of randomly oriented cylinders in space has been proposed (**Figure 7A**) to describe the observed non mono-exponential echo signal attenuation at high q/b values (35, 52, 128).
Assuming that the cylinder has a radius *a*, and the intracellular metabolites diffusivity is $D_{free}$, the measured echo signal represents the sum of signals from a large number of differently oriented cylinders. For any given cylinder whose axis makes a variable angle θ with the diffusion gradient, the echo signal attenuation is described by Eq. [5]. The total echo signal attenuation, S, as measured from a large spectroscopy voxel is then given by:

$$\frac{S(q,t_d)}{S(q=0,t_d)} = \frac{\int_0^{\pi/2} p(\theta)\frac{S^{\theta}_{cylinder}(q,t_d)}{S(q=0,t_d)}d\theta}{\int_0^{\pi/2} p(\theta)d\theta} \qquad [6]$$

where $p(\theta) = \sin(\theta)$ is the distribution of fibers making an angle $\theta$ relative to a fixed gradient direction in the case of an isotropic fiber distribution. If the fibers are expected to follow a specific orientation distribution (as in the case of a voxel entirely within white matter tracts) or a dispersion pattern, as for example in the DW-MRI NODDI model (129), Eq. [6] can be opportunely modified in order to take into account the specific fiber orientation or dispersion. While fiber orientation and dispersion is important and has to be taken into account, for example, in DW-MRS experiments in human brain white matter (see (130) and following **Section 3**), within the large spectroscopy voxel in preclinical experiments on mouse brain, fibers can be assumed isotropically distributed with good approximation (see Supplementary Information in (128)).
A thorough investigation of NAA diffusion in brain cells at high q/b values was done by Assaf and Cohen in the late 1990s (87, 90, 91). In these pioneering works (for a comprehensive review of early diffusion MRS studies we direct the reader to (29)), Assaf and Cohen characterized the restricted diffusion behaviour of NAA by showing its bi- and tri-exponential diffusion decays within a large range of b-values (up to 35000 s/mm$^2$) and diffusion times (up to 300 ms). Later on, Kroenke et al. and Yablonskiy and Sukstanskii (35, 52) proposed a first attempt to model NAA diffusion taking into account cellular structure by proposing a model of randomly oriented cylinders, in which the radial diffusivity was set to 0 (i.e., randomly oriented sticks). In this case, the total echo signal decay differs from the one in Eq. [6], and is described by:

$$\frac{S(q,t_d)}{S(q=0,t_d)} = \left(\frac{\pi}{4ADC_{axial}q^2t_d}\right)^{1/2} erf\left[(ADC_{axial}q^2t_d)^{1/2}\right] \qquad [7]$$

where $ADC_{axial}$ is the metabolite axial diffusivity, the $ADC_{radial}$ is set to 0, and erf[…] is the error function.

The choice to set the NAA radial diffusivity to 0 was based on the observation that the estimated radial diffusivity was an order of magnitude lower than the estimated axial one. However, the zero radial diffusivity assumption should not be considered true in systems with large nerve fibers and at short diffusion times, as shown, for example, by DT-MRS experiments on frog sciatic nerve (86). In fact, in this kind of systems, Ellegood and coworkers showed a non-zero radial diffusivity for NAA, tCr, tCho, Taurine (Tau) and Glutamine-Glutamate-GABA complex (Glx) (e.g. 0.15-0.18 μm$^2$/ms), for diffusion time of 18 ms (86).

Moreover, recent DW-MRS studies of other metabolites like Glu, tCr, tCho and Ins, in mouse brain showed that the assumption of zero-radius is not generally valid (128). Palombo et al. recently used the model in Eq. [9] to characterize healthy *in vivo* mouse brain at 11.7 T, showing that randomly oriented cylinders assumption accounts well for measured echo attenuation for Glu, tCr, Tau, tCho and Ins (**Figure 7B**), yielding fiber radii and $D_{free}$ in the expected ranges (0.5–1.5 μm and 0.30–0.45 μm$^2$/ms, respectively, **Figure 7C**) (128). Interestingly, the only exception was NAA, for which the extracted radial diffusivity and radius was 0. A small correction was proposed to the model and showed that the echo signal attenuation for NAA is compatible with a model where the majority of the NAA volume fraction diffuses in randomly oriented cylinders of non-zero radius, and a small fraction of the NAA (≤10%) is instead confined in highly restricted compartments where the NAA does not diffuse and has a short $T_2$ compared to the cytosolic NAA (128). The authors suggested that this small fraction may be representative of the NAA confined in mitochondria, where NAA is synthesized, and within the myelin sheath of neuronal axons. The introduction of this immobile NAA pool allowed the estimation of reasonable values for $D_{free}$ and fiber radius (~0.34 $\mu m^2$/ms and ~0.6 $\mu$m, respectively), supporting the effectiveness of the model in Eq. [6]. This fiber radius estimated from NAA diffusion was close to that estimated from Glu diffusion (~0.8 µm), while radii estimated from other metabolites appeared larger (the largest, ~1.6 µm, being found for Ins), suggesting that these non-neuronal metabolites are experiencing less radial diffusion in glial cells.

It may be surprising that DW-MRS allows the estimation of so small fiber diameters (<3 $\mu$m). Indeed, it is now well known that for water based DW-MRI, there is a lower limit to the sensitivity to fiber diameter which sets the minimum accessible diameter by using a single diffusion encoding sequence, $d_{min}^{(SDE)}$, at (131):

$$d_{min}^{(SDE)} = \left(\frac{768}{7}\frac{\sigma D_{free}}{\gamma^2 \delta g^2}\right)^{1/4} \qquad [8]$$

where $\sigma$ is the normalized standard deviation of the signal due to noise. For example, for the typical clinical high-SNR case (SNR = 50), where $\sigma$=1%, water $D_{free}$=2 $\mu m^2$/ms, g=80 mT/m and $\delta$=40 ms, $d_{min}^{(SDE)}$=3.3 $\mu$m. In similar experimental conditions, for metabolites with an average $D_{free}$=0.4 $\mu m^2$/ms, it is possible to reach $d_{min}^{(SDE)}$=2.2 $\mu$m. This significant improvement in diameter sensitivity is due to the much lower metabolites diffusivity compared to that of water molecules. Considering the specific preclinical experimental setup used in the high q/b values experiments reported in (97), for NAA with $D_{free}$=0.34 $\mu m^2$/ms, $\sigma$=1%, $g_{max}$=750 mT/m and $\delta$=3 ms, the resulting $d_{min}^{(SDE)}$ is 1.3 $\mu$m.

Experimental results and theoretical investigations reported in this session show very reasonable estimates for the brain cell fiber diameter. However, concerning fiber diameter estimation, one needs to be careful because there is still lack of experimental validations by direct histological measurements.

*Estimating fiber diameter from DDE RE MRS experiments*

Interestingly, the fiber diameter question for NAA and for Ins, representing the neuronal and astrocytic compartments of the CNS tissue, was tackled also from the DDE angle. The data in (126), other than providing evidence for fitting a randomly oriented anisotropic compartment model, can also be used for fitting the compartment's diameter. To do so, a large multidimensional dictionary of signals was simulated in the MISST toolbox (132, 133), with all sequence parameters (diffusion times, mixing times, gradient durations, gradient amplitudes, and number of measurements) input directly to the simulation. Based on the shape of the curve, a randomly oriented infinite cylinder model was chosen, with finely sampled diameters, and, importantly, the "free" diffusivity of the metabolite was also varied on a very fine grid for each diameter value. This provided a "fitting plane" ($D_{free}$, d), where $D_{free}$ is free diffusivity and d the diameter, to which the data was regressed. When the DDE modulation curves shown in **Figure 6** are carefully quantified for NAA and Ins resonances, using this approach, the fiber diameter was found to be less than 1.3 µm for NAA, and between 2-4 µm for Ins, with the local minima approaching 0.1 µm for NAA and 3.1 µm for Ins. These results represent the average diameter of randomly oriented cylinders. These results are very much in line with those reported for the high b/q values experiments described above.

To summarize, the two independent measurements – SDE high b/q values and angular DDE – provided similar results and reinforce each other. The NAA diffuses in fibers somewhat smaller than the astrocytes, and in both cases, for the experimental designs chosen, the sequences seem to be much more sensitive to the randomly oriented neurites or astrocyctic processes, respectively.

*Cellular long-range microstructure: cell fiber segment length and number of embranchments*

At long diffusion times (>100 ms), while it is possible to discard the effect of cell fiber diameter, which can be assumed to be zero, as well as finer secondary structures and consequent ADC time-dependency, the branching of cell fibers comprising the neuronal dendritic trees or astrocytes processes cannot be neglected. The only study investigating this so far showed that cell fiber finite length and embranchments induce a specific ADC time dependence at (ultra-)long $t_d$ (92). In a modeling framework that treats fibers as mono-dimensional branching objects "embedded" in a three-dimensional space and uses massive Monte-Carlo simulations, Palombo et al. predicted the effect of different morphometric statistics (i.e. the number of successive embranchments $N_{branch}$ along each process, and the segment length $L_{segment}$ for a given segment comprising a cellular fiber process) on the measured ADC time dependence. This general model was used to analyze data acquired up to $t_d$=2 seconds in the healthy mouse (at 11.7 T) and macaque (at 7 T) brain *in vivo* (**Figure 8**). The proposed modeling framework consistently classified cellular compartments, strongly supporting the generally accepted preferential compartmentalization of Ins and tCho inside astrocytes and of Glu and NAA in neurons, whereas some other metabolites such as tCr and Tau seem to have no preferential compartmentalization. In addition, extracted cell morphologies, such as length of branch segments and number of embranchments, were qualitatively and quantitatively consistent with histological data (**Figure 9**), suggesting that the effect of cell fiber length and embranchments must be considered when modeling long $t_d$ data.

## 3. The use of mutual information from DWI and DW-MRS

The task of extracting unequivocal microstructural information from diffusion weighted MR experiments is challenging, both in the case of DW-MRI and that of DW-MRS. In DW-MRI it is the lack of compartmental and cellular specificity and inter-compartmental exchange that pose the major challenge, as the sensitivity is high enough to allow reasonably low partial volume across white matter tracts, as well as enough "pure" gray and white matter voxels for tissue specific analysis. Conversely, in DW-MRS the major challenge stems from the intrinsic low sensitivity of the method, while the specificity, at least for some of the most prominent metabolites, is high. A natural consequence would thus be combining DW-MRI and DW-MRS in an analysis framework that benefits from the complementarity of the two methods. Few works so far have attempted to propose such a framework, and here we will briefly survey what has been done so far and what are some of the possible directions in which co-analysis of DW-MRI and DW-MRS data can evolve to provide a more comprehensive picture of tissue microstructure.

### 3.1 How can DW-MRS inform DW-MRI?

Insights from separate DW-MRS measurements that can have a deep impact on e.g. modeling of DW-MRI data have been already previously mentioned, e.g. the results from DW-MRS experiments in both animals and in humans that show no evidence for the existence of fully restricted compartments in neural tissue, up to diffusion times in the order of 1-2 seconds (48, 111). Another DW-MRS result with implications on the way DW-MRI data should be interpreted is that the ADC values of *all* metabolites in gray matter are consistently lower than in white matter (i.e. see **Table 1** or (134-136)) as opposed to the ADC of water, which is homogeneous across brain tissue when measured in the typical range of b≈1000 s/mm$^2$, and becomes higher in GM for higher values of b (137). This may indicate that either the extracellular space in cortical gray matter is more loosely packed than in white matter, or/and that there are intrinsic differences in the viscosity of these tissues, or/and that cross-membrane water exchange between the intra- and extracellular space is faster. The latter hypothesis obtains some support from recent experiments, including a measurement of apparent exchange rate (AXR) using the filter exchange imaging (FEXI) technique (138). So far, few studies provide separate metabolite ADC data for gray and white matter, and more evidence is needed to fully characterize metabolite ADC in white matter and cortical/subcortical gray matter, both in humans and in animals. A recent DW-MRS study in multiple sclerosis found that in the thalamus, for example, the ADC of NAA is *higher* than in parietal white matter (139), indicating yet another possible set of intracellular conditions that lead to higher diffusivity in that particular region.

To envision the potential benefits of *simultaneous* use of mutual information from DW-MRI and DW-MRS, it is useful to look at an early example of DW-MRS in animal model of stroke (80). In this work, as well as in subsequent works (50) a decrease in all metabolite ADC was measured in the acute ischemic phase. This is an independent evidence that supports the notion that the decrease in ADC in acute stroke is not solely a result of increased restriction of the extracellular space following intracellular edema (140). Additional support to these findings was provided by diffusion measurements of metabolites and water in brain slices (141) as well as in humans (83).

Another example is that examines diffusion tensor metrics of the intraneuronal metabolite, NAA, and those of water in the same two volumes of interest in the human corpus callosum (CC) (84). From the microstructural investigation stand point, the CC offers the simplest test-case possible: a single, easily identifiable white matter tract with a known orientation, which is mostly left-right in its medial part and then curves towards superior cortical regions. This structural simplicity led to the first work that modeled the diffusion of NAA in human white matter (52) and to the first reported diffusion tensor of metabolites from white matter regions (62). In Upadhyay et al (84), the diffusion tensors of both water and NAA are estimated from a DW-MRS experiment in which the water was suppressed for the spectroscopic data and left untouched for the water data. An estimation of the partial volume effect of CSF was essential for a more accurate estimation of the water tensor, but is not necessary for the estimation of the NAA tensor, as there is no NAA in the CSF. The FA values reported for NAA from the two callosal volumes were 0.72 and 0.52 for the anterior and more posterior VOI, respectively, and those of water were 0.46 and 0.39. It should be emphasized that the FA of water as calculated from the DW-MRS VOI is significantly lower than the one that could be obtained from a DTI experiment in the medial part of the CC. This is due to the macroscopic curvature of the callosal fibers within the spectroscopic VOI, which in this case similarly affects the NAA and the water FA. The fact that the water FA values are lower is not surprising, since in addition to the microscopically *and* macroscopically highly anisotropic intra-axonal space unique to the NAA, water is also present in other cellular structures, e.g. glia, as well as in the extracellular space and in the myelin sheath. Assuming that the contribution of the latter is negligible at the long TE in which the experiments were performed (142), it is theoretically possible to estimate, based on prior estimates of the intra-axonal and extra-axonal volume fractions in the VOIs, the FA of the *extra-axonal* compartment. The structural properties of the extra-axonal space in white matter are key in any modeling framework for DW-MRI data, and it is thus important to assess the validity of assumptions regarding its contribution to anisotropy, as these may greatly differ. This is indeed the case for the modeling framework CHARMED (143) and NODDI (129), where in CHARMED, as well as in other modeling frameworks (113, 144) the extra-axonal contribution to FA is independent of the intra-axonal contribution, whereas in NODDI the two contributions to FA are interdependent. It is important to note that when calculating FA values from DW-MRS experiments, noise propagation has to be carefully assessed, as FA values can become artificially high as a result of noisy measurements (43).

Another question regarding compartmental contributions to anisotropy is what is the volume fraction of the macroscopically *isotropic* compartment in tissue. It is safe to assume that a significant contribution to isotropic diffusion in tissue comes from structures that are *on average* isotropic. In neural tissue these can be glial cells such as astrocytes and microglia, although oligodendrocytes and fibrous astrocytes tend to align themselves to the white matter scaffold (145) and thus probably preserving some degree of overall anisotropy. DW-MRS can be key in answering this question. In the first study that reported on tensor DW-MRS measurements, the overall fractional anisotropy of metabolites measured in VOIs that ranged between 10-12cc was similar across metabolites, also in white matter regions with one predominant white matter tract (62). A more recent study utilized the increased sensitivity of DW-MRS at 7T, and examined the diffusion of the three main metabolites in the corpus callosum in a small VOI (2cc) with b values up to 11,000 s/mm$^2$. In this work, the ratio between the diffusivity parallel and perpendicular to the callosal fibers within a DW-MRS VOI, $ADC_{par}/ADC_{perp}$, was almost twice as high for NAA compared to tCho and tCr (146). Based on this finding, assuming that in white matter NAA is exclusively contained in axons, and that these are the sole source of intracellular macroscopic FA in the volume, it is possible to estimate the fractional volume of the isotropic compartments that contain the majority of tCho and tCr, based on their values of $ADC_{par}/ADC_{perp}$ and their overall tissue concentration. Based on the assumptions made above, it was estimated that the glial fraction of tCho is 0.5 and that of tCr is 0.4 ((146)). Although this estimate does not take into account the volume fraction of axons and glia in white matter, it supports the notion that tCho is highly present in glia, more so than e.g. tCr. Stronger diffusion anisotropy of NAA compared to tCr and tCho was also reported in the bovine optic nerve for the so-called "fast ADC" derived from a biexponential analysis (87), whereas in a peripheral nerve DW-MRS measurement, the FA of NAA and Glu, both neuronal metabolites, resulted lower than that of tCr and tCho (86). These diverse results call for more DW-MRS studies in different systems – *in vivo* as well as *in vitro*. A similar ratio to the one reported in (146) of $ADC_{par}/ADC_{perp}$ for NAA in the corpus callosum was reported earlier (52). This work, however, did not report findings from other metabolites.

### 3.2 How can DW-MRI inform DW-MRS?

Keeping in mind that the main limitation of DW-MRS is its limited sensitivity, reflected in a significantly lower SNR compared to DW-MRI, it is inevitable that the spatial resolution of DW-MRS would be much lower. To compensate for 3-4 orders of magnitude in concentration for most of the proton resonances of interest in the MR spectrum, a concomitant increase in volume is needed, as shown in the example of **Figure 10**. In standard MRI scanners this necessitates VOI of a few milliliters, thus with single dimensions in the order of centimeters. As a consequence, volumes of moderate size in white matter will include a broad axonal angular distribution, stemming both from the orientation dispersion across fibers and from the macroscopic factors such as the curvature of the fibers propagating within the VOI and multiple white matter tracts passing through the VOI (**Figure 10**). In the case of e.g. arbitrary VOI position in parietal white matter this results in an almost uniform directional distribution, as confirmed by examining the angular distribution of the principal eigenvectors (135). Even when the VOI is significantly smaller and positioned on a single tract such as the CC, the contribution of the macroscopic curvature to the orientational distribution within the VOI is significant (130, 146). Since the curvature and shape of the CC within the VOI may significantly vary across subjects, this macroscopic angular distribution is a source of unwanted variance to DW-MRS measurements that can obfuscate e.g. differences between metabolite diffusion properties across subject populations in studies that examine the diffusivity of NAA as a marker for axonal degeneration (63, 147).

As seen in **Section 2.3**, Eq. [6], for the diffusion of NAA in white matter, a way to account for this confound is to model the data assuming diffusion in a set of cylinders with a given angular distribution with respect to the gradient direction. The angular distribution generated by the macroscopic curvature of the tract can be obtained from a separately performed DTI experiment, from which the set of $E_1$, the principal eigenvectors of the diffusion tensors of voxels within the spectroscopic VOI can be extracted (130). This process is illustrated in stream A of **Figure 11**.

In an additional step, it is possible to include an additional, *microscopic* distribution in convolution with the macroscopic one, to account for axonal orientation dispersion (stream B in **Figure 11**). This results in a two-parameter model that not only accounts for the macroscopic confound, but also delivers an estimate for the orientation dispersion, shown to realistically fit with estimates from histology.

When DW-MRS is co-analyzed with DWI data, it is crucial to take into account the fact that actual localization of the various metabolites in the VOI is different due to frequency-dependent displacement of the VOI. This calls for accurate estimation of the real localization for each metabolite, especially for sequences such as PRESS, where the low bandwidth of the refocusing pulses can cause significant displacement along their slab selection axes.

Non-biased measurements of e.g. microscopic anisotropy of neuronal and glial processes can be also achieved directly from DW-MRS without using DTI data. This can be done by either the use of DDE experiments, as discussed in **Section 1.4 and 2.3**, or by generating the so-called "powder average" of multidirectional DW-MRS data at multiple diffusion weighting values, assuming that the sole source of deviation from monoexponential decay is the orientation dispersion (148). These methods are not mutually exclusive to each other, and at this stage cross validation of methods and their assessment with respect to stability, reproducibility and time consumption is essential.

### 3.3 Clinical implications and future directions in DW-MRS methodology

*Multivoxel DW-MRS and DW-MRSI*

Work that has been done so far in disease implies that not only DW-MRS is possible to apply in clinical settings, but that the potential of added information to that obtained from DWI experiments justifies the effort. So far, DW-MRS has been applied to study cerebral ischemia in animal models (50, 80, 81) and in humans (83, 149), cerebral tumors in animals (58) and humans (83), normal aging (149, 150), schizophrenia (147) and several studies in MS (63, 139, 151). Although valuable microstructural information can be obtained from combining simple DTI data and DW-MRS data acquired from a single volume, a significant effort should be invested in generating robust multivoxel DW-MRS data, as this will be essential for a spatially-resolved combination of data from the two modalities with significant gains for tissue microstructural characterization in healthy and more importantly for diseased conditions. As was previously mentioned, the cytosolic diffusion coefficient of NAA is a good candidate for a putative marker for intracellular damage such as axonopathy in MS, but this role can be extended to other neurodegenerative disorders, such as tauopathy in AD, where intraneuronal pathological changes are caused by hyperphosphorylation of the tau protein, which is responsible for keeping the integrity of microtubules (152-154). In a combined DTI – DW-MRS study of patients with neuropsychiatric systemic lupus erythematosus (NPSLE), an increase in the diffusion coefficient of tCho was observed in correlation with neuropsychiatric symptoms and with the SLE disease activity index (SLEDAI), which scales with inflammatory state (155). This points to the possibility that tCho's ADC is modulated by inflammation and reflects glial reactivity in response to inflammation. Reactive glia are known to undergo cytomorphological changes during activation, in line with increase in the ADC of intraglial metabolites (156). In this study, as well as in the DW-MRS study in multiple sclerosis., the DW-MRS findings correlated with DTI changes, and thus can mutually contribute to a better explanation of the disease process. The ability to do so in a spatially encoded manner over a large field of view is thus particularly attractive, as it offers the possibility to study water and metabolite diffusion in normal and abnormal appearing tissue, and map deficits across regions that may or may not correlate with disease outcome, together with e.g. PET tracers specific to cellular pathology.

A few efforts in developing sequences for DW-MRS in spectroscopic imaging mode have been published, both using conventional sequential k-space coverage, as well as using echo-planar spectroscopic imaging (EPSI) (136, 157-159). Conventional acquisition of magnetic resonance spectroscopic imaging (MRSI) data is long and prone to errors that stem from phase and amplitude fluctuations. While phase variations can be accounted for in the post-processing stage, amplitude fluctuations and signal drop-outs should be dealt with prospectively. Approaches that use navigators, i.e. short time-domain signal acquired after the diffusion encoding but prior to the spatial encoding, have been proposed and implemented on clinical scanners (159, 160), but robustness and sensitivity to disease effects has yet to be demonstrated. Efforts to shorten acquisition time would thus benefit from the use of parallel imaging (161), and if SNR is high enough, perhaps a degree of compressed sensing (162, 163).

DTI can also inform DW-MRS experiments in the planning phase, in cases where the diffusion properties of metabolites are to be studied e.g. along a particular white matter tract. In that case, DTI can be used to identify the tract of interest, and the DW-MRS can be subsequently performed on the desired tract, as was performed on the straight segment of the arcuate fasciculus, a tract involved in language processing (85). The selected volume can also be spatially encoded with the application of phase encoding gradient in one dimension (1D-MRSI). This method has been shown to yield the diffusion tensor of NAA in several voxels along the straight portion of the corpus callosum (164) in good geometric agreement with the diffusion tensors of water data from the same volumes.

Additional strategies have been proposed for multivoxel DW-MRS that may offer simple and robust solutions with faster acquisitions, at the expense of less than full coverage of the brain. One approach is based on simultaneous acquisition of e.g. two separate volumes, which can be separated in the post-processing stage using information about coil sensitivities (165, 166). This approach is similar to the multiband approach in imaging, e.g. (167). A dual volume acquisition approach can be useful in simultaneously acquiring data from two regions where one is visibly more affected by disease than the other, as in e.g. stroke or tumor that differentially affects one hemisphere.

## Conclusion

In this review article, we surveyed the state-of-the-art methods that have been developed for robust acquisition, quantification and analysis of DW-MRS data, and discussed the potential relevance of DW-MRS for elucidating brain microstructure *in vivo*. There is still much to be done to further develop DW-MRS and bring it to a broader audience, both in terms of acquisition methodology as well as data analysis and modelling. The ever-improving hardware – better and stronger gradient systems, more sensitive RF coils, and higher static magnetic fields – feeds the hope that DW-MRS will become increasingly more useful to the scientific community. Some encouraging examples were reported and discussed, showing that with accurate data on diffusion of increasing number of metabolites, and with accurate computational and geometrical modelling, metabolite DW-MRS can provide unique cell-specific information on the intracellular structure of brain tissue. Since the implementation of an imaging version of DW-MRS is still in its infancy, the integration of mutually compatible information derived from a combined DW-MRI and DW-MRS approach seems to be, at the moment, a more practicable route towards a better cell-specific characterization of brain microstructure.

## Acknowledgements

MP is grateful for funding from The Engineering and Physical Sciences Research Council (EPSRC) grant code: N018702. NS is grateful for funding from the European Research Council (ERC) under the European Union's Horizon 2020 research and innovation program (grant agreement No. 679058 - DIRECT-fMRI), as well as under the Marie Sklodowska-Curie grant agreement No 657366. JV is recipient of grant from the European Research Council about diffusion-weighted MRS (grant agreement No. 336331 - INCELL).

## Figures

| System | Sequence parameters | Metabolite | DT-MRS | | | | |
|---|---|---|---|---|---|---|---|
| | | | *$\lambda_1$ (μm²/ms)* | *$\lambda_2$ (μm²/ms)* | *$\lambda_3$ (μm²/ms)* | *MD (μm²/ms)* | *FA* |
| ***Human Brain WM+GM*** (Fotso et al., MRM 2016) | $B_0$=3 T; TE=90 ms $b_{max}$=1.7 ms/μm²; td=45 ms | NAA | // | // | // | 0.17 ± 0.04 | 0.57 ± 0.08 |
| | | tCr | // | // | // | 0.20 ± 0.04 | 0.63 ± 0.04 |
| | | tCho | // | // | // | 0.18 ± 0.04 | 0.57 ± 0.10 |
| ***Human Brain Occipital GM*** (Ellegood et al., NMR Biomed 2011) | $B_0$=3 T; TE=160 ms $b_{max}$=1.8 ms/μm²; td=75 ms | NAA | 0.35 ± 0.07 | 0.22 ± 0.05 | 0.10 ± 0.06 | 0.22 ± 0.05 | 0.53 ± 0.14 |
| | | tCr | 0.38 ± 0.07 | 0.22 ± 0.08 | 0.07 ± 0.02 | 0.22 ± 0.05 | 0.60 ± 0.10 |
| | | tCho | 0.35 ± 0.08 | 0.22 ± 0.07 | 0.09 ± 0.07 | 0.22 ± 0.05 | 0.54 ± 0.13 |
| ***Human Brain Subcortical WM*** (Ellegood et al., NMR Biomed 2011) | $B_0$=3 T; TE=160 ms $b_{max}$=1.8 ms/μm²; td=75 ms | NAA | 0.38 ± 0.06 | 0.24 ± 0.03 | 0.15 ± 0.04 | 0.26 ± 0.01 | 0.43 ± 0.13 |
| | | tCr | 0.39 ± 0.05 | 0.22 ± 0.02 | 0.12 ± 0.04 | 0.24 ± 0.02 | 0.51 ± 0.11 |
| | | tCho | 0.36 ± 0.07 | 0.20 ± 0.03 | 0.09 ± 0.04 | 0.22 ± 0.02 | 0.54 ± 0.15 |
| ***Human Brain Body Corpus Callosum*** (Ellegood et al., NMR Biomed 2011) | $B_0$=3 T; TE=160 ms $b_{max}$=1.8 ms/μm²; td=75 ms | NAA | 0.38 ± 0.01 | 0.24 ± 0.02 | 0.09 ± 0.02 | 0.23 ± 0.01 | 0.55 ± 0.06 |
| ***Human Brain Corticospinal Tract*** (Ellegood et al., NMR Biomed 2011) | $B_0$=3 T; TE=160 ms $b_{max}$=1.8 ms/μm²; td=75 ms | NAA | 0.51 ± 0.13 | 0.32 ± 0.05 | 0.08 ± 0.01 | 0.31 ± 0.06 | 0.52 ± 0.03 |
| ***Human Brain Left Arcuate Fasciculus*** (Upadhyay et al., NeuroImage 2008) | $B_0$=3 T; TE=135 ms $b_{max}$=1.7 ms/μm$^2$; td=50 ms | NAA | 0.27 ± 0.01 | $(\lambda_2+\lambda_3)/2$ = 0.12 ± 0.01 μm$^2$/ms | | 0.20 ± 0.01 | 0.53 ± 0.06 |
| ***Human Brain Right Arcuate Fasciculus*** (Upadhyay et al., NeuroImage 2008) | $B_0$=3 T; TE=135 ms $b_{max}$=1.7 ms/μm$^2$; td=50 ms | NAA | 0.25 ± 0.01 | $(\lambda_2+\lambda_3)/2$ = 0.14 ± 0.01 μm$^2$/ms | | 0.20 ± 0.01 | 0.43 ± 0.02 |
| ***Human Brain Body Corpus Callosum*** (Ronen et al., Brain Struct. Func. 2014) | $B_0$=7 T; TE=120 ms $b_{max}$=3.6 ms/μm$^2$; td=50 ms | NAA | 0.34 ± 0.03 | 0.24 ± 0.03 | 0.08 ± 0.01 | 0.22 ± 0.02 | 0.54 ± 0.04 |
| ***Human Brain Occipital GM*** (Ellegood et al., NMR Biomed 2011) | $B_0$=3 T; TE=160 ms $b_{max}$=5 ms/μm$^2$; td=75 ms | NAA | 0.23 ± 0.03 | 0.18 ± 0.03 | 0.14 ± 0.03 | 0.18 ± 0.02 | 0.25 ± 0.10 |
| | | tCr | 0.24 ± 0.03 | 0.19 ± 0.02 | 0.13 ± 0.03 | 0.19 ± 0.02 | 0.30 ± 0.07 |
| | | tCho | 0.21 ± 0.03 | 0.17 ± 0.02 | 0.12 ± 0.03 | 0.17 ± 0.03 | 0.28 ± 0.06 |
| ***Human Brain Subcortical WM*** (Ellegood et al., NMR Biomed 2011) | $B_0$=3 T; TE=160 ms $b_{max}$=5 ms/μm$^2$; td=75 ms | NAA | 0.33 ± 0.08 | 0.19 ± 0.03 | 0.12 ± 0.02 | 0.21 ± 0.02 | 0.47 ± 0.13 |
| | | tCr | 0.32 ± 0.07 | 0.19 ± 0.02 | 0.09 ± 0.02 | 0.20 ± 0.02 | 0.51 ± 0.13 |
| | | tCho | 0.26 ± 0.06 | 0.15 ± 0.05 | 0.08 ± 0.01 | 0.16 ± 0.02 | 0.51 ± 0.14 |
| ***Human Brain Body Corpus Callosum*** (Ellegood et al., NMR Biomed 2011) | $B_0$=3 T; TE=160 ms $b_{max}$=5 ms/μm$^2$; td=75 ms | NAA | 0.34 ± 0.01 | 0.19 ± 0.00 | 0.08 ± 0.02 | 0.21 ± 0.01 | 0.56 ± 0.03 |
| ***Human Brain Corticospinal Tract*** (Ellegood et al., NMR Biomed 2011) | $B_0$=3 T; TE=160 ms $b_{max}$=5 ms/μm$^2$; td=75 ms | NAA | 0.34 ± 0.04 | 0.22 ± 0.02 | 0.15 ± 0.11 | 0.24 ± 0.06 | 0.41 ± 0.19 |

**Table 1**: Summary of diffusion tensor (DT) MRS metrics (DT eigenvalues $\lambda_1$, $\lambda_2$ and $\lambda_3$, mean diffusivity, MD, and fractional anisotropy, FA) estimated in different white (WM) and gray (GM) matter regions in healthy human brain together with the corresponding references.

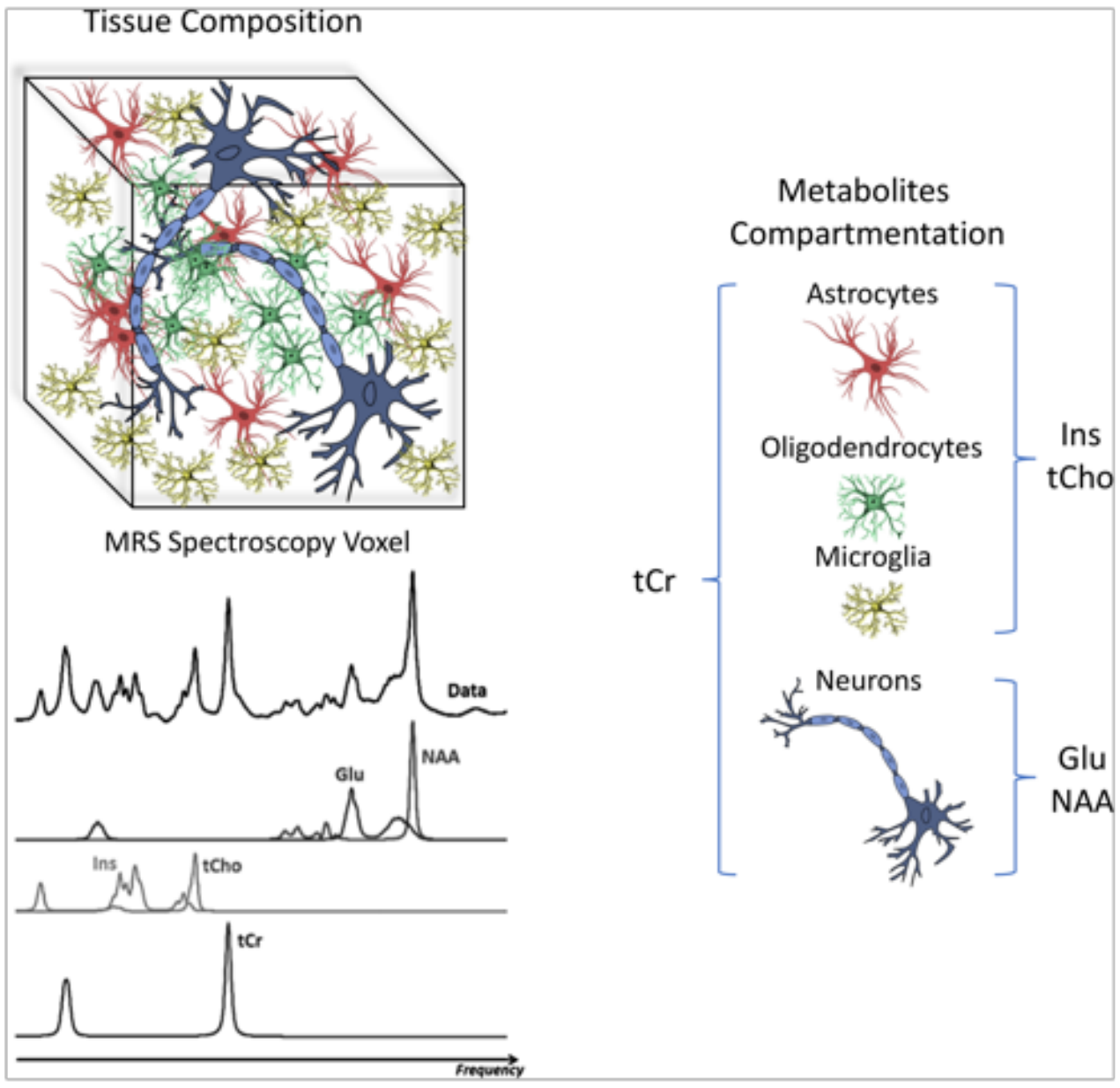

**Figure 1**: Schematic representation of the composition of a spectroscopic voxel within the brain tissue, together with the generally accepted (although still debated) cell-specific compartmentalization of brain metabolites. Typical MRS spectra and the specific peaks associated to the most investigated metabolites (NAA, Glu, tCr, tCho and Ins) are also schematized.

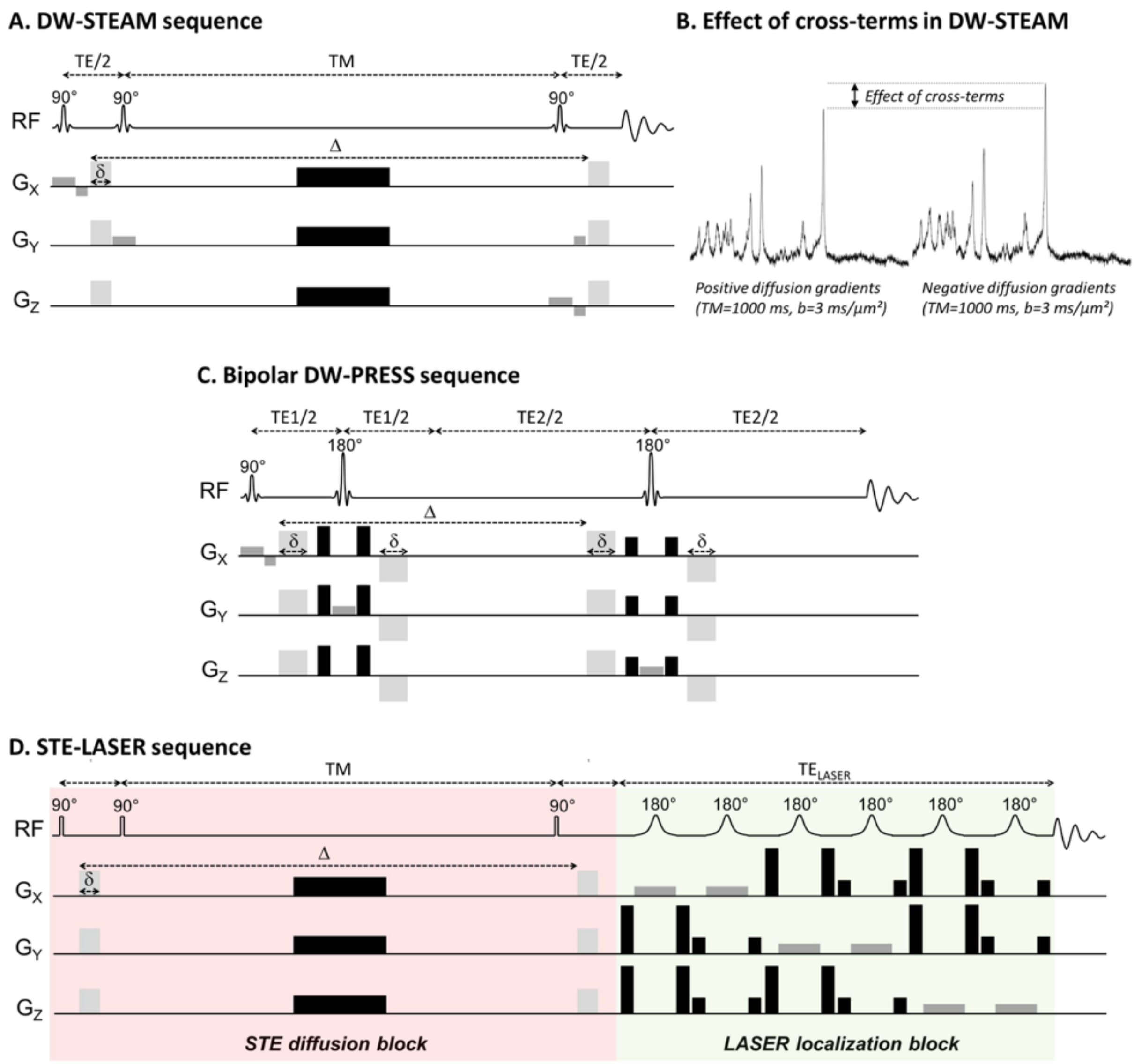

**Figure 2**: A) A diffusion-weighted stimulated echo acquisition mode (DW-STEAM) sequence built in such a way that cross-terms are minimized (X- and Z-slice refocusing is performed immediately after the first and third 90° pulses, and diffusion gradients are used as spoilers during the echo time). However, cross-terms cannot be totally suppressed, as Y-slice refocusing corresponding to the second 90° pulse must be performed during the second part of the echo, after the third 90° pulse, i.e. diffusion and Y-slice selection gradients are not refocused during the whole mixing time, which might result in large cross-term at long TM / low b value. B) Example of the manifestation of cross-terms on DW-spectra acquired in the monkey brain at 7 T (taken from (48)). Signal attenuation is different when diffusion gradient polarity is positive and negative, which is due to cross-terms, whose sign is changed with gradient polarity. C) A bipolar DW-PRESS sequence allowing reaching relatively large b-values including in Humans (52), while keeping eddy currents and cross-terms low. D) The stimulated echo localization by adiabatic slice refocusing (STE-LASER) sequence (60) where the stimulated echo diffusion block precedes the LASER localization block, avoiding any cross-term due to overlap between dephasings induced by diffusion and localization gradients. Diffusion gradients are in light grey, slice selection gradients in dark grey, and spoiler gradients in black.

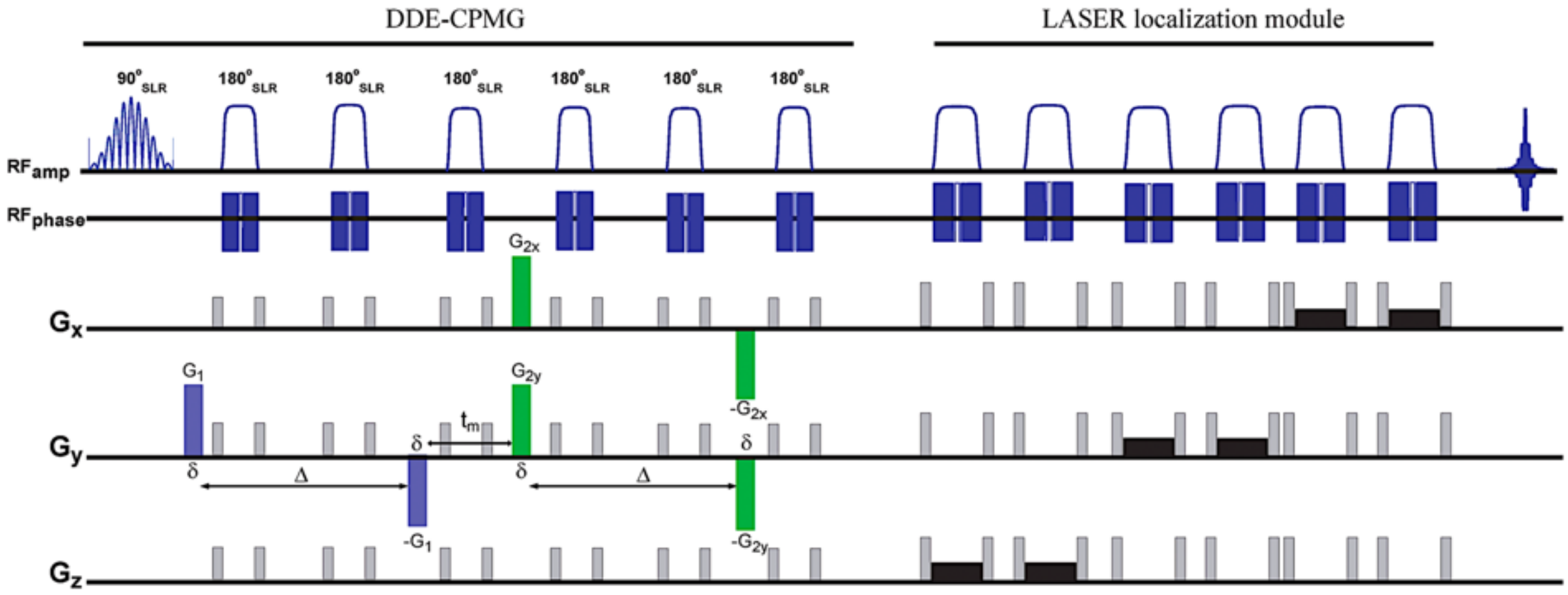

**Figure 3**: General scheme for relaxation enhanced (RE) MRS. The sequence begins with a spectrally-selective excitation, followed by a filter, and then localization using localization by adiabatic slice refocusing (LASER); the signal is finally acquired either in spectroscopy or spectroscopic imaging mode. In this case, the filter proposed was a double diffusion encoding (DDE) embedded within a Carr-Purcell-Meiboom-Gill (CPMG) block (spectrally selective refocusing pulses), to mitigate cross-terms with susceptibility-driven, internal gradients. In general, note that RE MRS aims to avoid manipulation of the water signal such that it remains at its original equilibrium position ($M_z$), thereby obviating the need for water suppression.

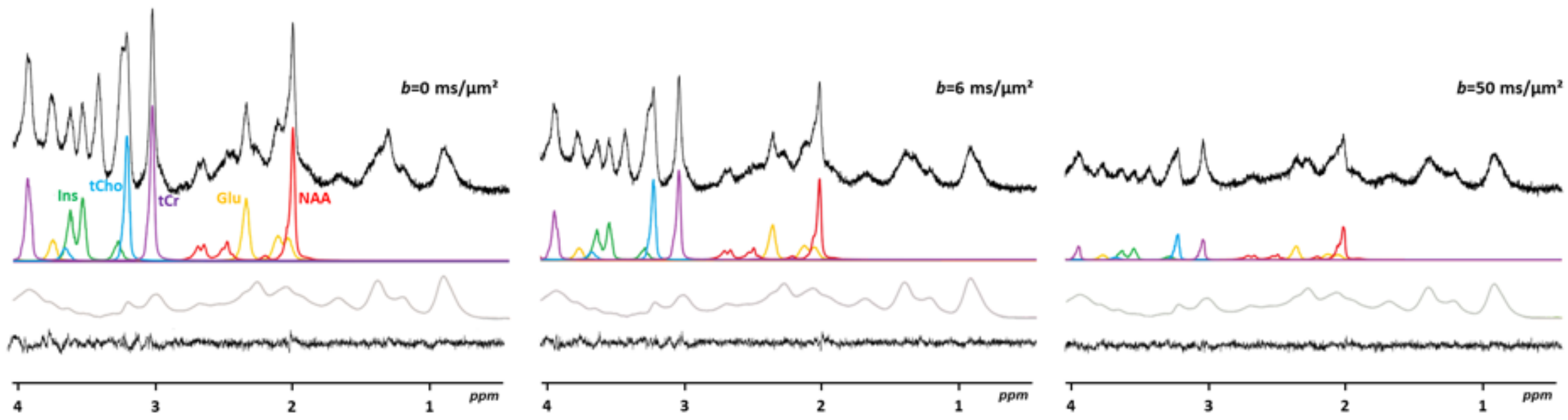

**Figure 4**: Examples of linear combination analysis of DW-spectra acquired in the mouse brain at 11.7 T using the stimulated echo localization by adiabatic slice refocusing (STE-LASER) sequence (TE/TM=33.4/50 ms), for three different b-values. A subset of five important metabolites, as fitted by LCModel, is shown on the first row immediately below the raw data. On the row below, the macro molecules (MM) contribution (experimental spectrum acquired with metabolite-nulling and incorporated into LCModel's basis-set) is shown. It is obvious that MM contribution cannot be ignored, in particular at high b-values. The row just above the ppm scale is the fit residual, which ideally should be a flat line with Gaussian noise if the modeled spectrum can perfectly explain the data.

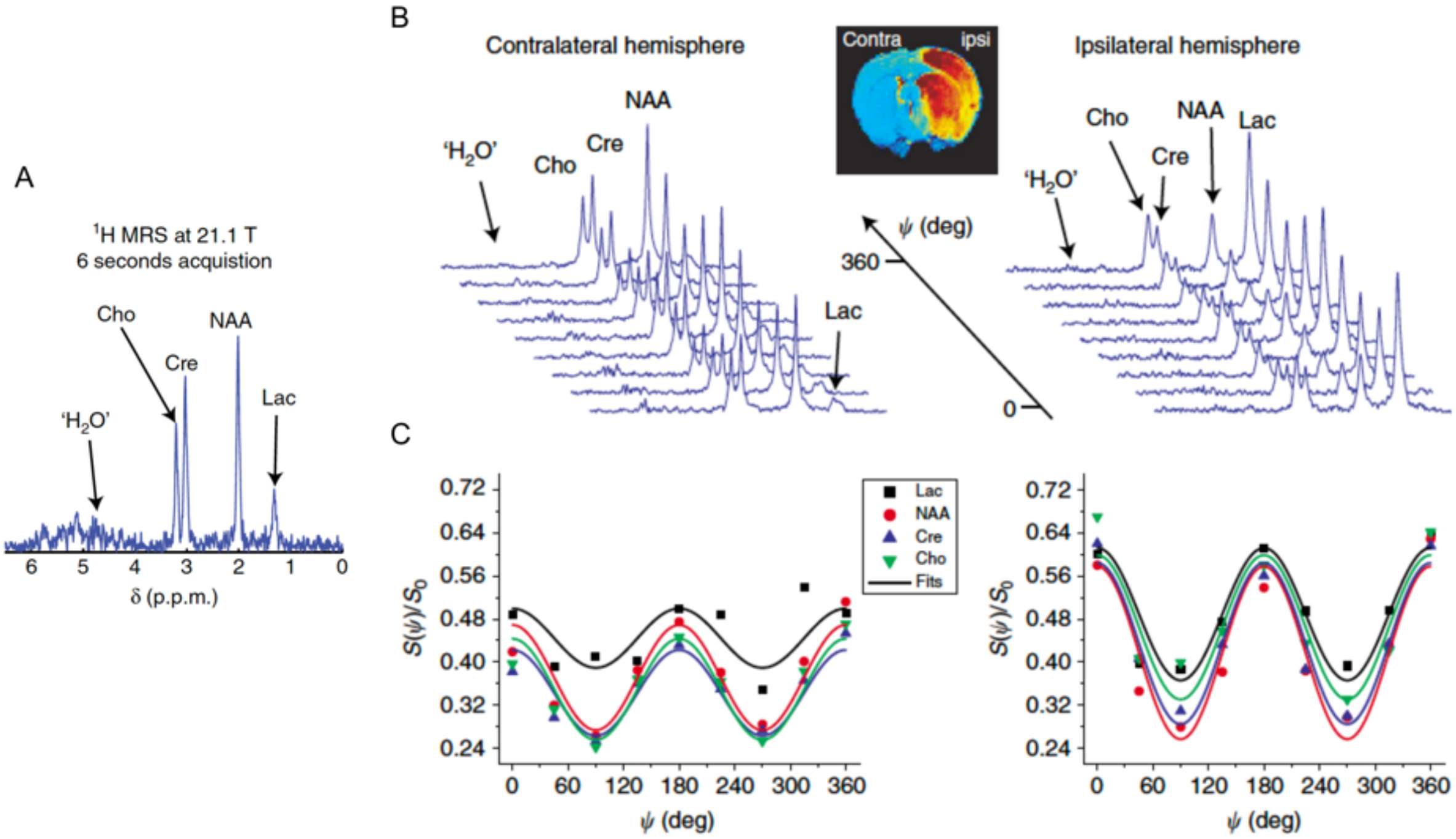

**Figure 5**: A) Relaxation enhanced (RE) MRS spectra (no diffusion weighting) acquired in only 6 seconds from a 5x5x5 (mm)$^3$ volume in the *in vivo* rat brain at 21.1T. Pulses were designed such that only the NAA, Lac, tCr (Cre) and tCho (Cho) peaks were excited/refocused; no water suppression was used. High fidelity spectra were observed, with SNR of the NAA peak exceeding 50, and only minor residual water appearing in the spectrum, suggesting <0.1% excitation around the water resonance. B) Raw data (stacked plots) of angular double diffusion encoding (DDE) RE MRS experiments performed on a similar voxel placed in the contra- (left panel) or ipsi-lateral (right panel) of a representative stroked rat 24 h post ischemia. The following parameters were used: gradient amplitudes of 48 G/cm, equal

diffusion times of 53.5 ms, diffusion gradient durations of 2.5 ms and a mixing time of 24 ms, at TR/TE = 1500/187 ms. The spectra show excellent quality, and the DDE modulation can be observed with the naked eye for both hemispheres; noticed the pronounced Lac signal emerging in the affected area. C) DDE modulation curves extracted for the different metabolites, evidencing both the restricted diffusion in randomly oriented compartments and microstructural differences between ipsi/contra-lateral hemispheres after ischemia. For details, see (59).

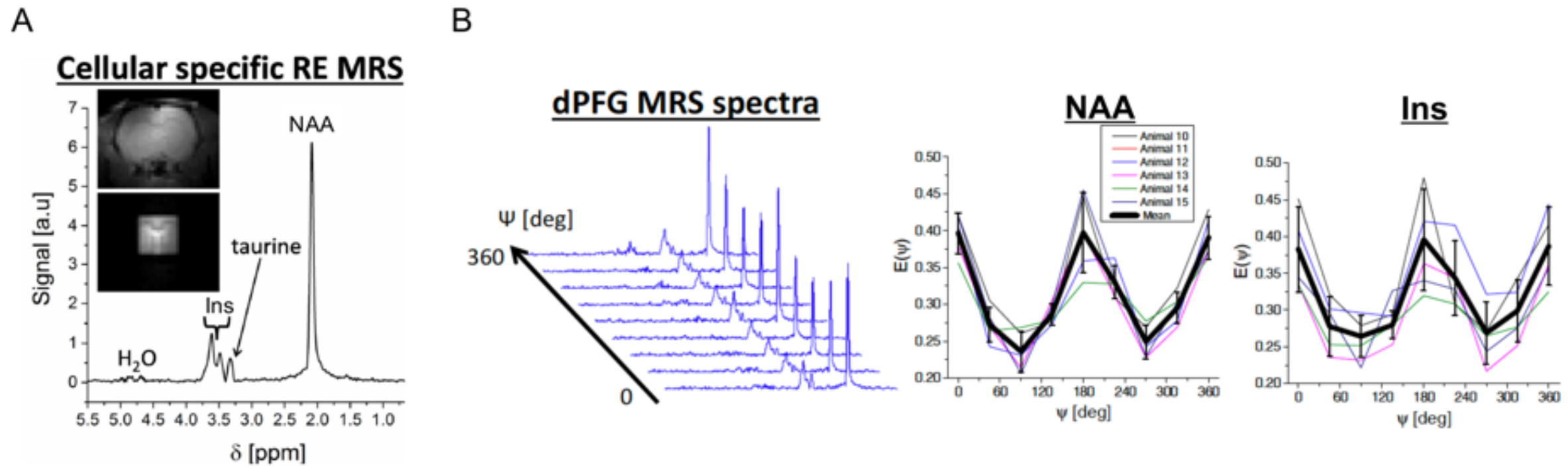

**Figure 6**: A) Relaxation enhanced (RE) MRS voxel (inset shows the localization) and spectra for a non-diffusion-weighted sequence seeking to isolate the NAA and Ins resonances. This clean spectrum shows RE MRS's ability to quantify even typically challenging Ins signals; importantly, the spectrum bears a signature for neurons and astrocytes from the NAA and Ins signals, respectively. B) The raw double diffusion encoding (DDE) RE MRS data (left panel in (B)) and ensuing DDE modulation curves (right panel in (B), thin lines represent individual animals and the thick lines are the mean). Experimental parameters were similar to those given in the caption of **Figure 4** (though not exactly identical), n.b., the experiments were performed on healthy animals. The DDE modulation is again observed with the naked eye even in raw data, and shows rather low variability. The NAA and Ins clearly diffuse in randomly oriented compartments characterized by restricted diffusion. Note that, considering the actual composition of the RE MRS voxel (~20% white matter; ~60% gray matter; ~20% CSF), the randomly oriented fibers model is a valid approximation, according to (128). Quantification of the modulation shows that NAA diffuses in extremely small cylinders with diameters <0.1 µm, while Ins diffuses in cylinders with diameter ~3 µm (126).

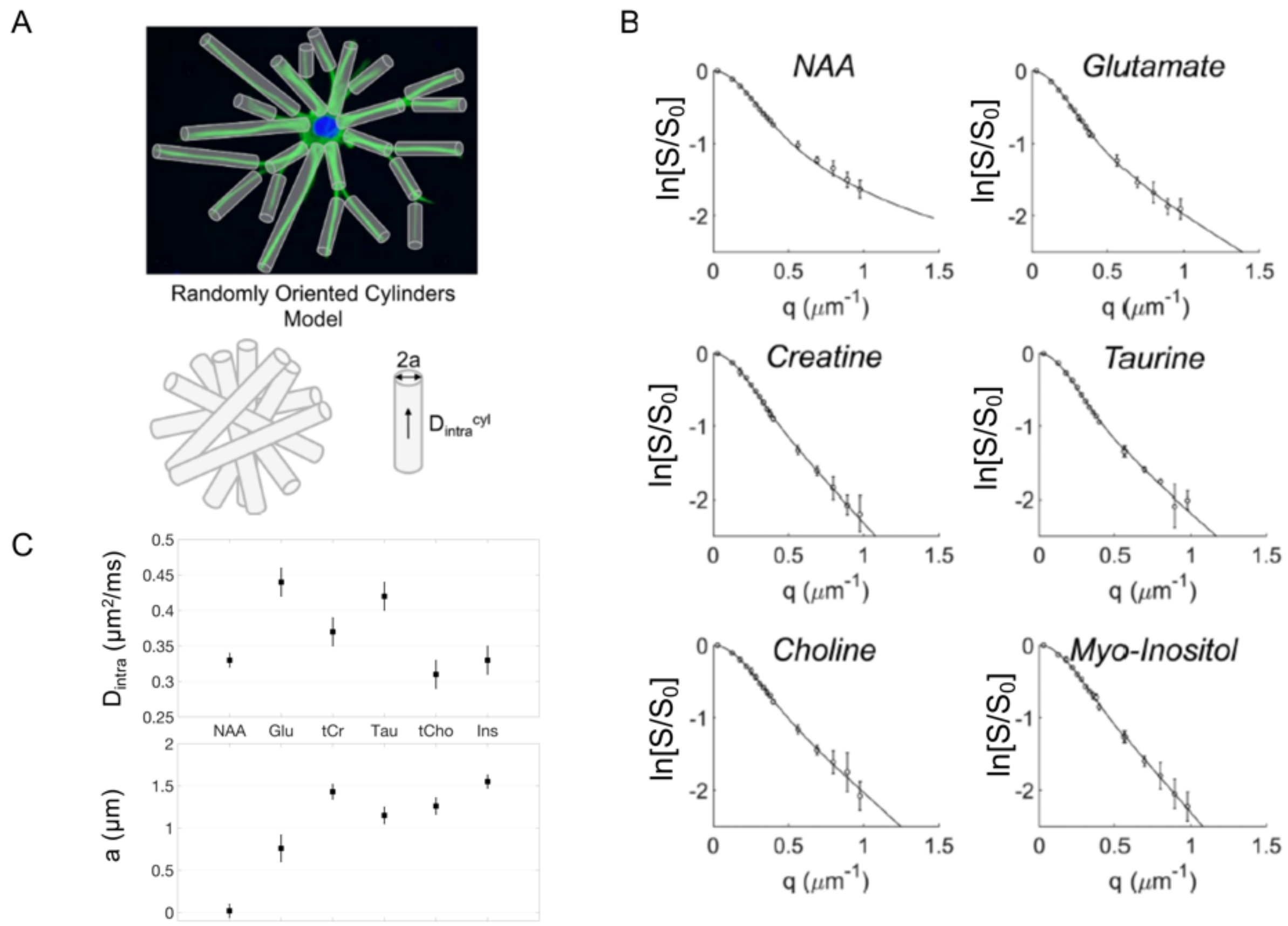

**Figure 7**: A) Schematic description of the randomly oriented cylinders model used to fit the experimental data. B) DW signal attenuation (points) and corresponding fitted curves (lines) as a function of q for all of the investigated metabolites. Error bars denote the SD. C) Estimated model parameters from the fit of the randomly oriented cylinders model (Eq. [6]) to experimental data of each metabolite. Note: $D_{intra}$ = intracellular diffusivity; *a* = cylinder's radius (mean ± SD, 2500 Monte Carlo draws). From (128).

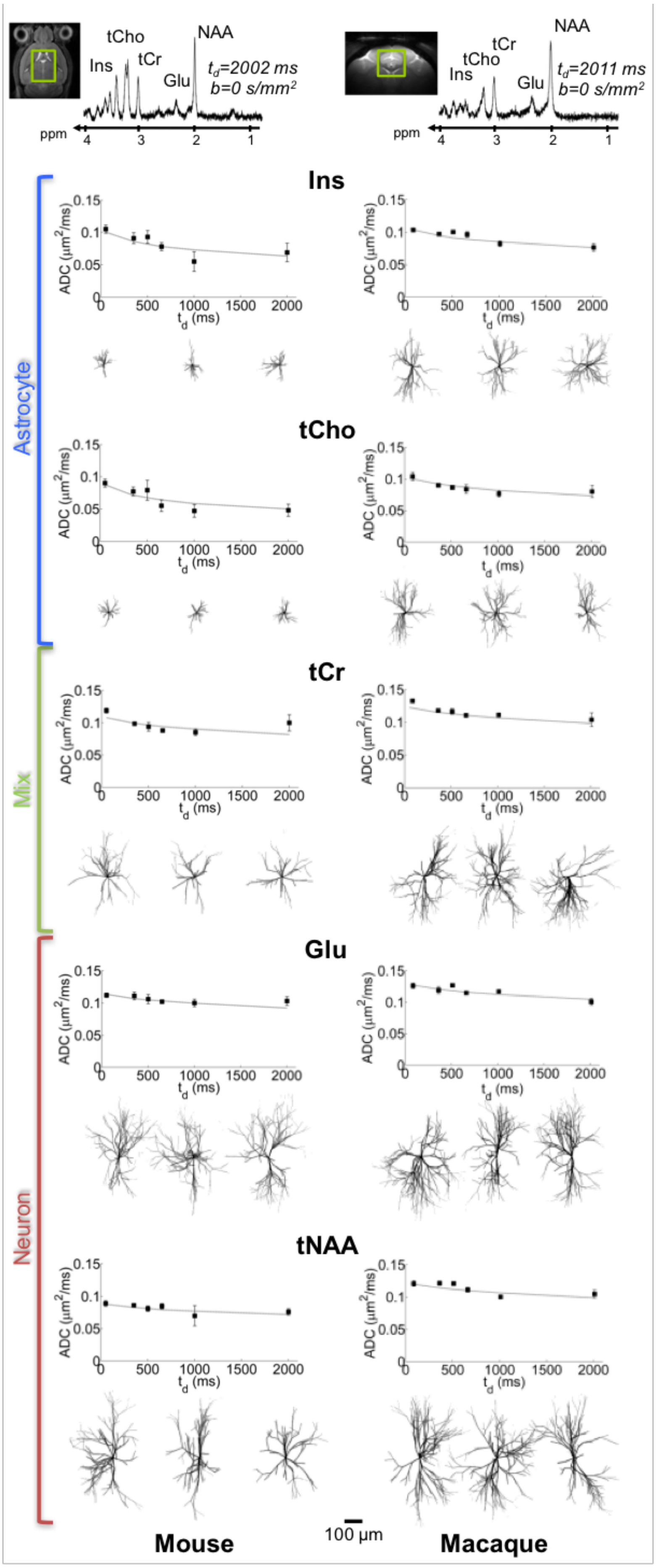

**Figure 8**: DW-MRS results and modeling in the mouse and macaque brain. The investigated volume of interest within the brain (green box) and a typical DW-MRS spectrum at $t_d$ = 2 s (and b=0 s/mm$^2$), as used to measure (using diffusion sensitizing gradients simultaneously applied along the three main axis x, y and z) ADC time dependence for each metabolite (Inset plots), are shown for each species. Points and error bars stand for ADC means and standard deviation of the means, respectively, estimated among the cohorts. Best fit of ADC (averaged over the cohorts) is also displayed as a continuous curve. A subset of the extracted synthetic cells for each metabolite is also reported. (Scale bar, 100 μm.). From (92). Note that, considering the actual composition of the RE MRS voxel (~20% white matter; ~60% gray matter; ~20% CSF), the randomly oriented fibers model is a valid approximation, according to (128).

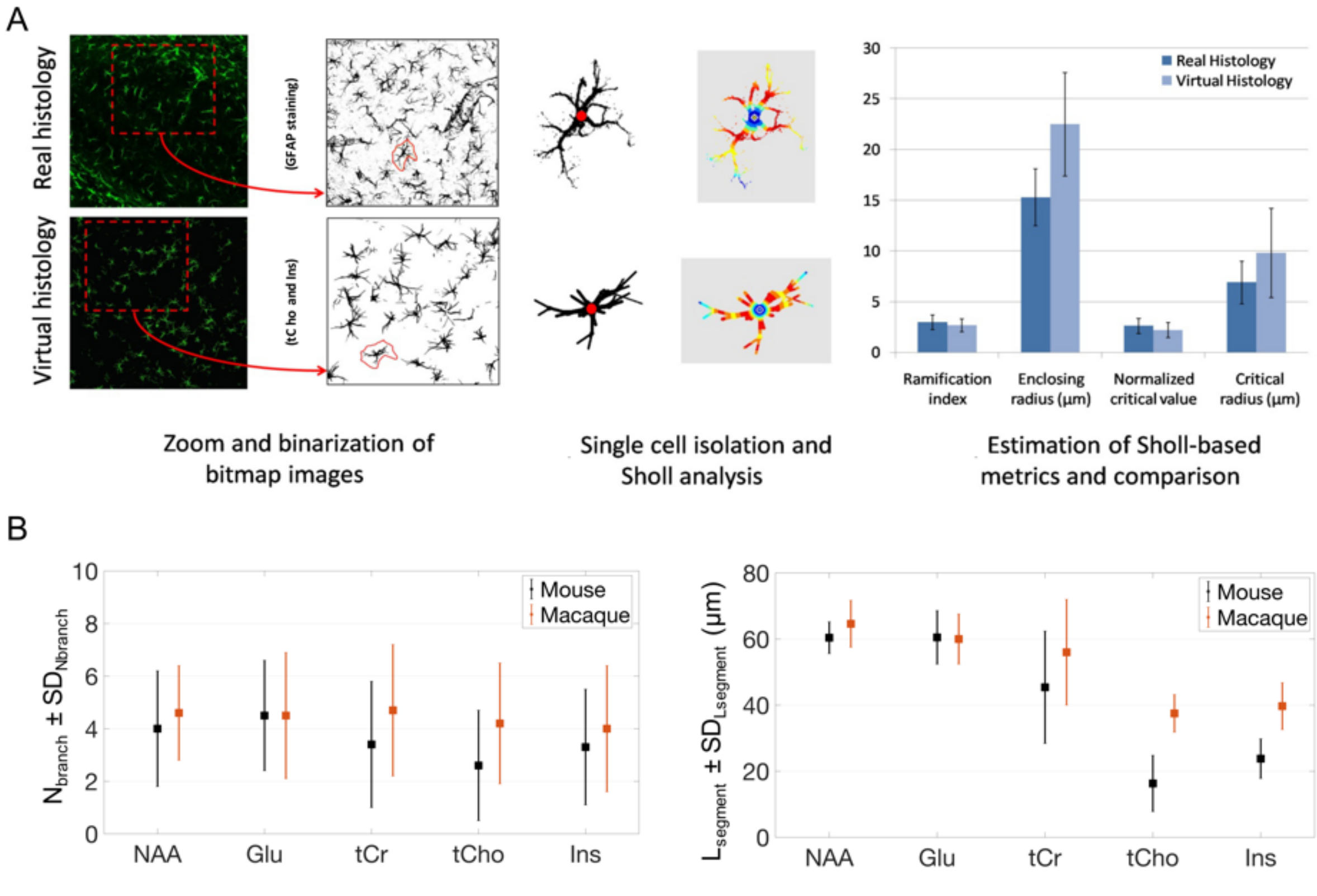

**Figure 9**: A) Examples of the Sholl analysis on a single astrocytic cell from a real glial fibrillary acidic protein (GFAP) stained hippocampal slice of mouse brain and a virtually reproduced one. Starting from left to rigth, each cell is isolated within the investigated real and synthetic histological slice; its center is identified; and the Sholl analysis, based on statistics from concentric circles (drawn in different colors), is performed. Sholl analysis results from 135 different cells were taken into account to estimate the mean and s.d. of Sholl-based metrics and here reported as histograms. No statistically significant differences were found between the Sholl-based metrics measured from real and virtual histological slices. B) Morphometric parameters estimated for the metabolite compartments by fitting the ADC time dependency in the mouse and macaque brain (see **Figure 6**). Metabolites thought to be preferentially compartmentalized in astrocytes are indicated by the letter A, those thought to be preferentially compartmentalized in neurons are indicated by the letter N, and those thought to be evenly mixed are indicated by A+N. From (92).

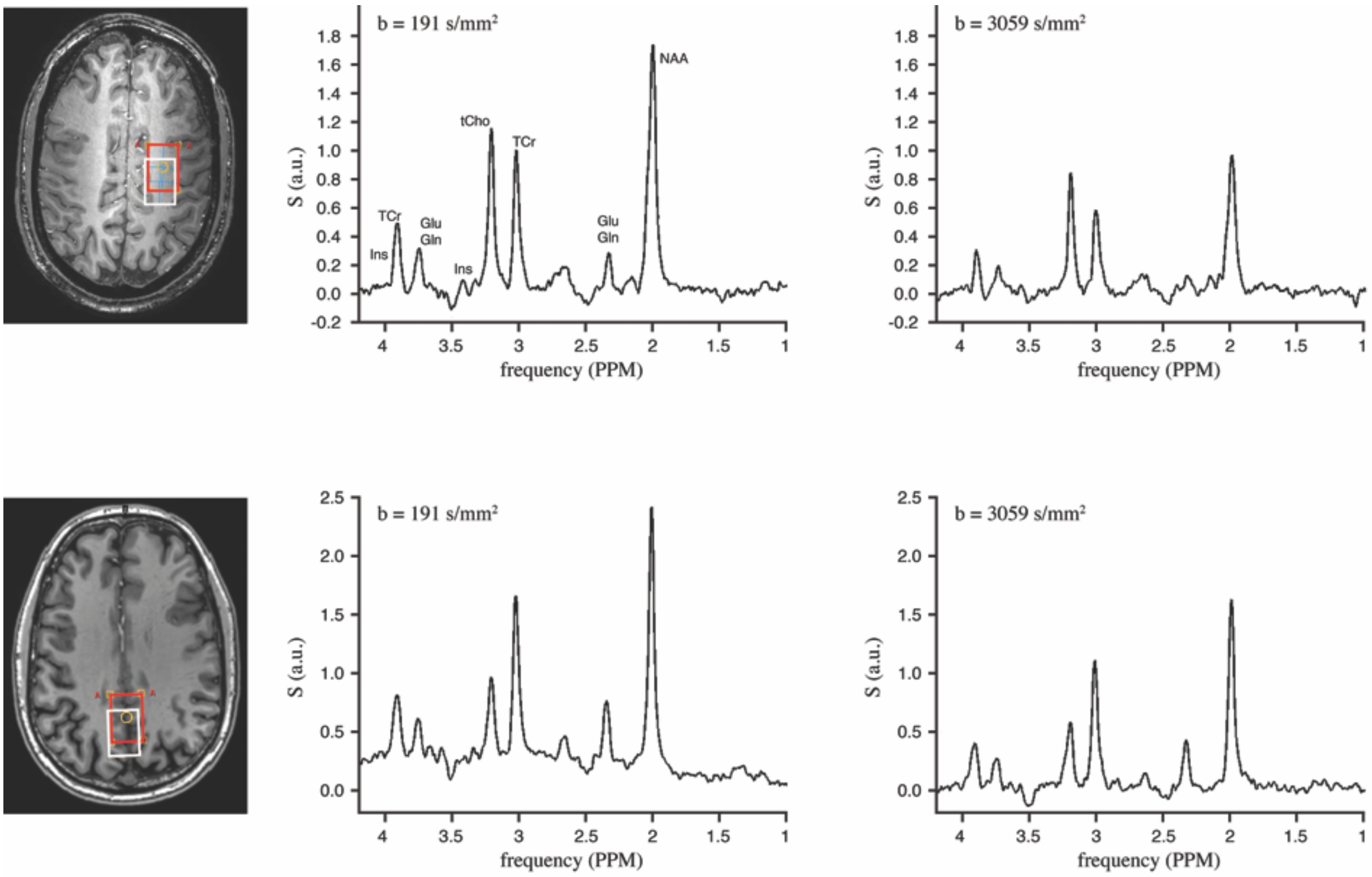

**Figure 10**: Voxel locations in the parietal white matter and occipital grey matter and the corresponding spectra at low and high b-value. Voxel locations are shown for NAA at 2.01 ppm (red box) and myo-inositol at 4.05 ppm (white box), the latter is shifted due to chemical shift displacement. Voxel locations are shown in white matter (top) and grey matter (bottom). On the right, the corresponding spectra are shown for b = 191 s/mm$^2$ and b = 3059 s/mm$^2$ with gradient direction [1,0,1]. In the grey matter spectra, the reduction in signal intensity for instance NAA and Glu and Gln is clearly reduced compared with white matter.

Peak assignments: NAA = N-acetyl aspartate, tCr = creatine and phosphocreatine, tCho = choline containing compounds, Ins = myo-inositol, Glu = glutamate, and Gln = glutamine. Spectra are shown with 2 Hz line broadening.

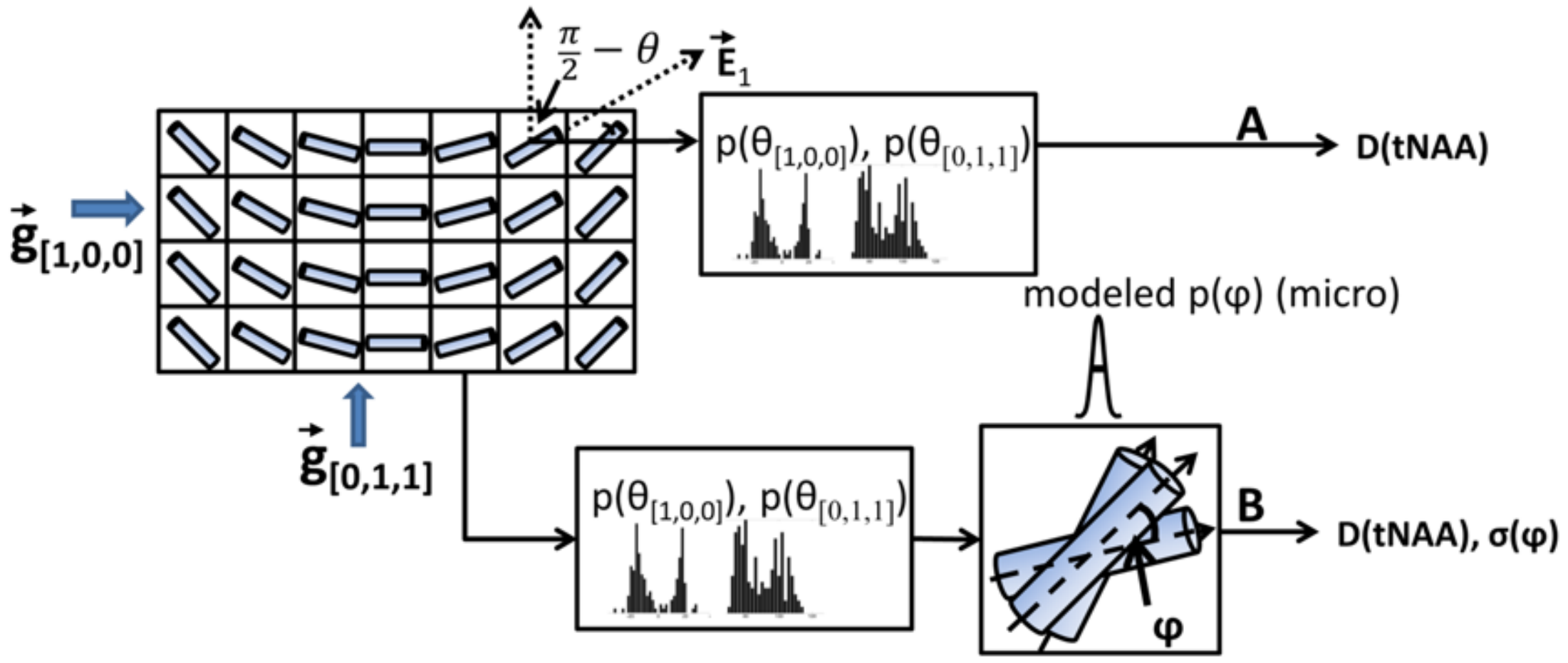

**Figure 11**: Schematic diagram of two possible modeled analyses of tNAA DWS data that use DTI information within the spectroscopic volume. In (A) experimental distributions $p(\theta_{[0,1,1]})$ and $p(\theta_{[1,0,0]})$ are estimated to account for macroscopic curvature of the corpus callosum within the volume of interest (VOI). These distributions are derived from information about $E_1$, the main eigenvector of a DTI data set estimated for all DTI voxels that lie within the spectroscopic VOI. The only fitted parameter is D(tNAA). (B) An additional residual angular distribution, $p(\phi)$, is introduced based on microstructural factors such as fiber orientation dispersion. In this analysis path, the standard deviation of this distribution, $\sigma_\phi$, is estimated from the data fitting procedure in addition to D(tNAA). From (130).